\pdfoutput=0
\documentclass[11pt]{article}
\usepackage{axodraw}
\usepackage{epsfig}
\usepackage{amsfonts}
\usepackage{amsmath}
\usepackage{bm,bbm}
\usepackage{cite}
\allowdisplaybreaks[1]

\input pix.sty

\newcommand{\kallen}{\sqrt{\lambda}}
\newcommand{\diag}{\mathop{\mbox{diag}}}
\newcommand{\K}{\rmii{$K$}}

\newcommand{\T}{\rmii{$T$}}

\newcommand{\nS}{n_\rmii{$S$}}
\newcommand{\nG}{n_\rmii{$G$}}

\newcommand{\aL}{a^{ }_\rmii{L}}
\newcommand{\aR}{a^{ }_\rmii{R}}
\renewcommand{\eq}{eq.~}
\renewcommand{\eqs}{eqs.~}
\renewcommand{\se}{sec.~}
\renewcommand{\ses}{secs.~}
\renewcommand{\fig}{fig.~}
\renewcommand{\figs}{figs.~}

\newcommand{\alphas}{\alpha_{\rm s}}
\newcommand{\Nf}{N_{\rm f}}
\newcommand{\Nc}{N_{\rm c}}

\newcommand{\mE}{m_\rmii{E}}

\newcommand{\gammaE}{\gamma_\rmii{E}}
\newcommand{\rmO}{{\mathcal{O}}}

\newcommand{\CF}{C^{ }_3} % {C^{ }_\rmii{F}}
\def\lsi{\raise0.3ex\hbox{$<$\kern-0.75em\raise-1.1ex\hbox{$\sim$}}}
\def\gsi{\raise0.3ex\hbox{$>$\kern-0.75em\raise-1.1ex\hbox{$\sim$}}}

\newcommand{\sign}{\mathop{\mbox{sign}}}
\newcommand{\nF}{n_\rmii{F}}
\newcommand{\nB}{n_\rmii{B}}
\newcommand{\rmii}[1]{{\mbox{\tiny\rm{#1}}}}
\newcommand{\rmiii}[1]{{\mbox{\tiny{$\scriptstyle{\rm#1}$}}}}

\newcommand{\im}{\mathop{\mbox{Im}}}
\newcommand{\Tint}[1]{{\hbox{$\sum$}\!\!\!\!\!\!\!\int\,}_{\!\!\!\!\raise-0.9ex\hbox{$\scriptstyle{#1}$}}}
\newcommand{\Tinti}[1]{{{\Sigma}\!\!\!\!\raise0.3ex\hbox{$\int$}_\rmii{${#1}$}}}

\newcommand{\bi}{\begin{itemize}}
\newcommand{\ei}{\end{itemize}}
\newcommand{\hide}[1]{ }
\newcommand{\bsl}[1]{\,\slash\!\!\!\!{#1}\,}
\newcommand{\msl}[1]{\,\slash\!\!\!{#1}\,}
\newcommand{\rsl}[1]{\,\slash\hspace*{-2.2mm}{#1}\hspace*{0.1mm}}
\newcommand{\scat}[1]{\mbox{scat}^{ }_\rmi{$#1$}}
\newcommand{\s}[1]{s_{#1}}

\newcommand{\E}{\mathcal{E}}

\renewcommand{\P}{\mathcal{P}}

\renewcommand{\K}{\mathcal{K}}
\newcommand{\MM}{M^2}

\newcommand{\mH}{m_\rmii{$H$}} % {m_\rmiii{Z}}
\newcommand{\ala}{{\ell}} % {{\ell_a}}

\newcommand{\aS}{\phi} % {{S}}

\newcommand{\aQ}{\gamma} % {Q}
\newcommand{\ag}{g}

\newcommand{\bla}{{\tilde\ell}}

\newcommand{\cS}{\phi} % {\rmiii{S}}
\def\TAsc(#1,#2)(#3,#4,#5)%
{\SetWidth{2.0}\CArc(#1,#2)(#3,#4,#5)\SetWidth{1.0}}
\def\Lwidth{3}

\def\TAgl(#1,#2)(#3,#4,#5){\SetWidth{2.0}\PhotonArc(#1,#2)(#3,#4,#5){\Lwidth}%
{6.283 #3 mul 360 div #4 #5 sub #4 #5 sub mul sqrt mul Tdensity mul}%
\SetWidth{1.0}}
\def\TLgl(#1,#2)(#3,#4){\SetWidth{2.0}\Photon(#1,#2)(#3,#4){\Lwidth}
{#1 #3 sub #1 #3 sub mul #2 #4 sub #2 #4 sub mul add sqrt Tdensity mul}%
\SetWidth{1.0}}
\renewcommand{\picc}[1]{\;\parbox[c]{60pt}{\begin{picture}(60,30)(0,-10)
\SetWidth{1.0}\SetScale{1.3} #1 \end{picture}}\; }
\def\Lwidth{1.3}

\renewcommand{\pic}[1]{\;\parbox[c]{30pt}{\begin{picture}(30,30)(0,-3)
\SetWidth{1.0}\SetScale{0.8} #1 \end{picture}}\;}
\renewcommand{\picb}[1]{\;\parbox[c]{45pt}{\begin{picture}(45,30)(0,-3)
\SetWidth{1.0}\SetScale{0.8} #1 \end{picture}}\;}
\makeatletter \@addtoreset{equation}{section} \makeatother
\renewcommand{\theequation}{\arabic{section}.\arabic{equation}}
\makeatletter
\renewcommand\section{\@startsection {section}{1}{\z@}%
                                   {-5.5ex \@plus -1ex \@minus -.2ex}% bfr-
                                   {2.3ex \@plus.2ex}%
                                   {\normalfont\large\bfseries}}
\renewcommand\subsection{\@startsection{subsection}{2}{\z@}%
                                     {-3.25ex\@plus -1ex \@minus -.2ex}%
                                     {1.5ex \@plus .2ex}%
                                     {\normalfont\normalsize\bfseries}}
\renewcommand\thesection {\@arabic\c@section}
\renewcommand\thesubsection   {\thesection.\@arabic\c@subsection}
\renewcommand{\@seccntformat}[1]{%
\csname the#1\endcsname.\hspace{1.0em}}
\makeatother
\begin{document}

\flushbottom

\begin{titlepage}

\begin{flushright}
% OUTLINE  \\ 
% DRAFT \\ 
% arXiv:1008.3263\\ 
October 2021
\end{flushright}
\begin{centering}
\vfill

{\Large{\bf
 Smooth interpolation between thermal Born and LPM rates
}} 

\vspace{0.8cm}

J.~Ghiglieri$^\rmi{a}$
 and 
M.~Laine$^\rmi{b,c}$
 
\vspace{0.8cm}

$^\rmi{a}$%
{\em
SUBATECH, Universit\'e de Nantes, IMT Atlantique, IN2P3/CNRS,\\
4 rue Alfred Kastler, La Chantrerie BP 20722, 44307 Nantes, France \\}

\vspace{0.3cm}

$^\rmi{b}$%
{\em
AEC, 
Institute for Theoretical Physics, 
University of Bern, \\ 
Sidlerstrasse 5, CH-3012 Bern, Switzerland \\}

\vspace{0.3cm}

$^\rmi{c}$%
{\em
Department of Physics, 
P.O.Box 35 (YFL), \\ 
FI-40014 University of Jyv\"askyl\"a, Finland \\}

\vspace*{0.8cm}

\mbox{\bf Abstract}
 
\end{centering}

\vspace*{0.3cm}
 
\noindent
In a weakly coupled ultrarelativistic plasma,
% with particle masses small compared with the temperature,
$1 + n \leftrightarrow 2 + n$ scatterings,
with $n \ge 0$, need sometimes to be summed to all orders, 
in order to determine a leading-order interaction rate.
To implement this ``LPM resummation'', 
kinematic approximations are invoked. 
However, in cosmological settings, 
where the temperature changes by many orders of magnitude  
and both small and large momenta may play a role, such  
approximations are not always justified. We suggest a procedure to 
smoothly interpolate between LPM-resummed $1 + n \leftrightarrow 2 + n$
and Born-level $1 \leftrightarrow 2$ results, rendering the outcome
applicable to a broader range of masses and momenta. The procedure is 
illustrated for right-handed neutrino production from a Standard Model
plasma, and dilepton production from a QCD plasma.  

\vfill

%% %\noindent
%% %PACS numbers: 
%% %11.10.Wx, %        Finite temperature field theory
%% %11.15.Ha, %        Lattice gauge theory  
%% %12.38.Bx, %        Perturbative calculations in QCD
%% %12.38.Mh, %        Quark--gluon plasma
%% %14.40.Nd, %        Bottom mesons
%% %\\
%% %Keywords: Thermal Field Theory, CP violation, Neutrino Physics, Resummation
 
\end{titlepage}

\tableofcontents

%%%%%%%%%%%%%%%%%%%%%%%%%%% SECTION %%%%%%%%%%%%%%%%%%%%%%%%%%%%%%%%%%%%%%
%
\section{Introduction}

Many problems in particle cosmology and heavy ion collision physics
involve the computation of thermal interaction rates. On one hand this
comes about in the context of production rates, for instance of 
``freeze-in'' dark matter candidates, gravitational waves, 
right-handed neutrinos influencing leptogenesis, or photons and 
dileptons in the collider setting. On the other hand
interaction rates dictate how efficiently probes approach
equilibrium or keep up with it, as is relevant for instance 
for ``freeze-out'' dark matter candidates, active neutrinos, 
or for the quenching of an energetic jet produced in an initial hard 
scattering or decay. 

In terms of Feynman diagrams, the lowest-order topologies yielding
an interaction rate are $1\leftrightarrow 2$ processes, 
dubbed decays and inverse decays. 
However, the phase space available for these reactions 
is strongly constrained, implying that the corresponding interaction
rate is proportional to a positive power of particle masses. 
If we go to the so-called ultrarelativistic (UR) regime, 
where all masses are small compared with the temperature $T$, then 
the rate of $1\leftrightarrow 2$ processes gets suppressed
compared with the scaling dimension $T$.

If a process is phase-space suppressed, it can experience a large
correction by a soft additional scattering, which modifies the kinematics. 
Therefore $1+n\leftrightarrow 2+n$ processes, with $n\ge 0$, 
need to be considered, and in some cases summed to all orders. The
corresponding physics is related to that governing the 
propagation of high-energy cosmic rays through the atmosphere, 
whereby the treatment goes 
under the name of Landau-Pomeranchuk-Migdal (LPM) 
resummation~(cf.,\ e.g.,\ refs.~\cite{gelis3,amy1,agmz,bb1} and
references therein).
An incomplete list of recent cosmological applications of LPM resummation
can be found in 
refs.~\cite{bb1,app1,interpolation,broken,app2,degenerate,app3,app4,app5,app6}.

It is typical of resummations that 
their consistent implementation requires the presence of a scale 
hierarchy. In the case of LPM resummation, the scale hierarchy 
is that characterizing the UR regime, i.e.\ with masses small 
compared with momenta, the latter of which are of order~$\pi T$. 
However, in cosmology, the temperature changes, 
so a given particle can be ultrarelativistic at early and 
non-relativistic at late times. Moreover, we are often
interested in the overall abundance of a certain species, 
so that all momenta are integrated over. In these situations, 
the scale hierarchy justifying the LPM resummation gets 
compromised. In previous studies, the issue has been 
circumvented by somewhat {\it ad hoc} interpolations 
between LPM-resummed and Born-level
computations~\cite{broken,degenerate}. Recipes for switching off 
LPM resummation for large virtualities have been introduced for 
the dilepton case as well~\cite{nlo_dilepton,harvey}.  
The goal of the present paper is to suggest 
how the interpolation can be implemented
on the integrand level, rendering it smooth and numerically robust. 

We start by working out the kinematics of Born-level 
${1}\leftrightarrow{2}$ processes in a special coordinate system
which permits us to put the expression in a form similar 
to that appearing in LPM resummation (cf.\ \se\ref{se:1to2}). 
After recalling typical implementations of LPM resummation 
(cf.\ \se\ref{se:LPM}), it is then straightforward to 
suggest an interpolation (cf.\ \se\ref{se:interpolation}). 
As consistency checks,
we discuss how the result compares with well-known ultraviolet (UV) 
asymptotics at large virtualities (cf.\ \se\ref{se:OPE}),  
and verify that it correctly 
accounts for infrared (IR) divergences  
(cf.\ \se\ref{se:subtraction}). After illustrating 
the interpolation numerically (cf.\ \se\ref{se:numerics}), 
we summarize the recipe, 
and suggest how it can be extended to include specific 
NLO corrections (cf.\ \se\ref{se:concl}).

%%%%%%%%%%%%%%%%%%%%%%%%%%% SECTION %%%%%%%%%%%%%%%%%%%%%%%%%%%%%%%%%%%%%%
%
\section{Phase space for thermal $1\leftrightarrow 2$ processes}
\la{se:1to2}

%%%%%%%%%%%%%%%%%%%%%%%%%%% SUBSECTION %%%%%%%%%%%%%%%%%%%%%%%%%%%%%%%%%%%
%
\subsection{General derivation in light-cone coordinates}

We start by recalling the derivation of the phase space
average for thermal $1\leftrightarrow 2$ reactions at the Born
level, arriving at an expression (cf.\ \eq\nr{1to2_step2})
which can subsequently be interpolated
(cf.\ \se\ref{se:interpolation})
to the result obtained from LPM resummation 
(cf.\ \se\ref{se:LPM}). 
The derivation may look complicated and 
requires a few opportune choices of variables, 
but this is necessary for establishing \eq\nr{1to2_step2} 
in its desired form.  

For determining the
interaction rate originating from $1\leftrightarrow 2$ reactions 
at the Born level 
(this is generically denoted by $\Gamma^\rmi{Born}_{1\leftrightarrow 2}$,
remarking that an overall normalization factor is needed for obtaining
the physical rate, cf.\ footnotes~\ref{rhn} and \ref{dilepton}), 
it is sufficient to compute the functional form 
originating from $1\to 2$ decays of a would-be 
non-equilibrium particle. 
The four-momentum of the non-equilibrium particle 
is denoted by $\K \equiv (\omega,\vec{k})$.  
The corresponding matrix
element squared, or ``splitting function'', 
summed over the spins and degeneracies of
the final-state particles, reads $\Theta(\P^{ }_a,\P^{ }_b)$.
Here $\P^{ }_{a,b}$ are the four-momenta of particles of types~$a$
and~$b$, both of which are assumed thermalized.
Then the full rate reads % (for $\omega > 0$) 
\ba
 \Gamma^\rmi{Born}_{1\leftrightarrow 2}
 & = &  
   \scat{1\to 2}(a^{ },b^{ })
 \, \Theta(\P^{ }_a,\P^{ }_b)
 \nn 
 & + &  
   \scat{2\to1}(-a^{ };b^{ })
 \, \Theta(-\P^{ }_a,\P^{ }_b)
 \la{ex2} \\ 
 & + &
    \scat{2\to1}(-b^{ };a^{ })
 \, \Theta(\P^{ }_a,-\P^{ }_b)
 \;, \nonumber
\ea
where we have defined thermal phase space averages as 
\ba
 \scat{1\to2}(a,b) & \equiv & 
 \frac{1}{2} 
    \int \! {\rm d}\Omega^{ }_{1\to2} 
          \, \mathcal{N}^{ }_{a,b} 
 \;, \la{scat1to2} \\ 
%%%%
 {\rm d}\Omega^{ }_{1\to2} & \equiv & 
 \frac{1}{(2\pi)^6}
 \frac{{\rm d}^3\vec{p}^{ }_a}{2 \epsilon^{ }_{a}}
 \frac{{\rm d}^3\vec{p}^{ }_b}{2 \epsilon^{ }_{b}}
 \, (2\pi)^4 \delta^{(4)}(\K - \P^{ }_a  - \P^{ }_b )
 \;, \la{Phi1to2} \\[2mm] 
%%%%
 \mathcal{N}^{ }_{a,b} 
 & \equiv & 
 \bar{n}^{ }_{\sigma_a}(\epsilon^{ }_{a} - \mu^{ }_a)
 \,\bar{n}^{ }_{\sigma_b}(\epsilon^{ }_{b} - \mu^{ }_b)
 - 
   n^{ }_{\sigma_a}(\epsilon^{ }_{a} - \mu^{ }_a)
 \,n^{ }_{\sigma_b}(\epsilon^{ }_{b} - \mu^{ }_b)
 \;,  \\[2mm]
%%%%%%%%%%%%%%%%%%%%%%%%%%%%%%%%%%%%%%%%%%%
 \scat{2\to1}(-a;b) & \equiv & 
 \frac{1}{2} \int \! {\rm d}\Omega^{ }_{2\to1} 
          \, \mathcal{N}^{ }_{a;b} 
 \;, \la{scat2to1} \\ 
%%%%
 {\rm d}\Omega^{ }_{2\to1} & \equiv & 
 \frac{1}{(2\pi)^6}
 \frac{{\rm d}^3\vec{p}^{ }_a}{2 \epsilon^{ }_{a}}
 \frac{{\rm d}^3\vec{p}^{ }_b}{2 \epsilon^{ }_{b}}
 \, (2\pi)^4 \delta^{(4)}(\K + \P^{ }_a   - \P^{ }_b )
 \;, \la{Phi2to1} \\[2mm] 
%%%%
 \mathcal{N}^{ }_{a;b} 
 & \equiv & 
 n^{ }_{\sigma_a}(\epsilon^{ }_{a} + \mu^{ }_a)
 \,\bar{n}^{ }_{\sigma_b}(\epsilon^{ }_{b} - \mu^{ }_b)
 - 
 \bar{n}^{ }_{\sigma_a}(\epsilon^{ }_{a} + \mu^{ }_a)
 \,n^{ }_{\sigma_b}(\epsilon^{ }_{b} - \mu^{ }_b)
 \;. \hspace*{5mm} 
%%%%%%%%%%%%%%%%%%%%%%%%%%%%%%%%%%%%%%%%%%%%%%%
\ea
Here $\sigma = +(-)$ labels bosons (fermions), and the
corresponding distribution functions are
\be
 \bar{n}^{ }_{\sigma}(\epsilon) \; \equiv \; 1 + n^{ }_{\sigma}(\epsilon) 
 \;, \quad
 n^{ }_{\sigma}(\epsilon) \; \equiv \; \frac{\sigma}{e^{\epsilon/T} - \sigma}
% \;, \quad
% \sigma = \pm
 \;, \la{n_sigma}
\ee
where $T$ is the temperature. 
The overall factor $\fr12$ in \eqs\nr{scat1to2} and \nr{scat2to1} 
is a convention, guaranteeing that $\Theta$ can be interpreted as 
the matrix element squared $\sum |\mathcal{M}|^2$ 
of a Boltzmann equation,\footnote{%
 In case of identical final-state particles, this should be replaced by 
 $
  \frac{1}{2} \sum |\mathcal{M}|^2
 $
 as usual. 
 } 
though a division by $\omega$ is still needed for obtaining the rate proper. 

Denoting the masses of the various particles by 
$M^2 \equiv \mathcal{K}^2 \equiv \omega^2 - k^2$, 
$m_a^2 \equiv \P_a^2$, 
the energy-momentum conservation constraints 
in \eqs\nr{Phi1to2} and \nr{Phi2to1} imply 
that the three channels in \eq\nr{ex2} 
get realized if 
\ba
 \epsilon^{ }_b & = & \omega - \epsilon^{ }_a
 \;, \quad 
 M > m^{ }_a + m^{ }_b 
 \;, \\ 
%%%%
 \epsilon^{ }_b & = & \omega + \epsilon^{ }_a
 \;, \quad 
 m^{ }_b > m^{ }_a + M  
 \;, \\ 
%%%%
 \epsilon^{ }_b & = & \epsilon^{ }_a - \omega 
 \;, \quad 
 m^{ }_a > m^{ }_b + M  
 \;, 
\ea
respectively. 
The goal is to integrate over $\vec{p}^{ }_b$
in \eqs\nr{scat1to2} and \nr{scat2to1}, and then to 
combine the channels into a single expression, 
with the remaining average given by 
an integral over~$\epsilon^{ }_a$.

As a first step, 
making use of 
$
 n(-x) = - \bar{n}(x)
$, 
indicating the energy arguments of $\mathcal{N}^{ }_{a,b}$ explicitly, 
and carrying out the integral over $\vec{p}^{ }_b$, we can rewrite
\eq\nr{ex2} as 
\ba
 & & \hspace*{-0.5cm}
 \Gamma^\rmi{Born}_{1\leftrightarrow 2}
  =  
 \int \! \frac{{\rm d}^3\vec{p}^{ }_a}{2(4\pi)^2} \, \biggl\{ 
 \nn
 & + & 
 \frac{\mathcal{N}^{ }_{a,b}(\epsilon^{ }_a - \mu^{ }_a, 
                             \omega - \epsilon^{ }_a - \mu^{ }_b)}
      {\epsilon^{ }_a \epsilon^{ }_{b(\vec{k} - \vec{p}_a)}}
      \bigl[
      \delta(\omega - \epsilon^{ }_a - \epsilon^{ }_{b(\vec{k} - \vec{p}_a)} ) 
      - 
      \delta(\omega - \epsilon^{ }_a + \epsilon^{ }_{b(\vec{k} - \vec{p}_a)} ) 
      \bigr]
      \, 
      \Theta(\P^{ }_a,\K - \P^{ }_a)
 \nn
 & + & 
 \frac{\mathcal{N}^{ }_{a,b}(- \epsilon^{ }_a - \mu^{ }_a, 
                             \omega + \epsilon^{ }_a - \mu^{ }_b)}
      {\epsilon^{ }_a \epsilon^{ }_{b(\vec{k} + \vec{p}_a)}}
      \bigl[
      - 
      \delta(\omega + \epsilon^{ }_a - \epsilon^{ }_{b(\vec{k} + \vec{p}_a)} ) 
      \bigr]
      \, 
      \Theta(-\P^{ }_a,\K + \P^{ }_a)
 \, \biggr\} 
 \;. 
\ea
The Dirac-$\delta$ constraints can be combined into
\ba
 & & \hspace*{-0.5cm}
 \Gamma^\rmi{Born}_{1\leftrightarrow 2}
  =  
 \int \! \frac{{\rm d}^3\vec{p}^{ }_a}{(4\pi)^2} \, \biggl\{ 
 \nn
 & + & 
 \frac{\mathcal{N}^{ }_{a,b}(\epsilon^{ }_a - \mu^{ }_a, 
                             \omega - \epsilon^{ }_a - \mu^{ }_b)}
      {\epsilon^{ }_a}
      \sign(\omega - \epsilon^{ }_a)
      \delta\bigl[
                   (\omega - \epsilon^{ }_a)^2
                 - \epsilon^{2}_{b(\vec{k} - \vec{p}_a)} 
            \bigr] 
      \, 
      \Theta(\P^{ }_a,\K - \P^{ }_a)
 \nn
 & - & 
 \frac{\mathcal{N}^{ }_{a,b}(- \epsilon^{ }_a - \mu^{ }_a, 
                             \omega + \epsilon^{ }_a - \mu^{ }_b)}
      {\epsilon^{ }_a }
      \sign(\omega + \epsilon^{ }_a)
      \delta\bigl[
                   (\omega + \epsilon^{ }_a)^2
                 - \epsilon^{2}_{b(\vec{k} + \vec{p}_a)} 
            \bigr] 
      \, 
      \Theta(-\P^{ }_a,\K + \P^{ }_a)
 \, \biggr\} 
 \;. \nn \la{ex5} 
\ea

To carry out the integral over $\vec{p}^{ }_a$, one normally 
goes over to spherical coordinates, selecting $\vec{k}$
as the $z$-axis. 
However, in order to make contact with 
LPM resummation~\cite{gelis3,amy1,agmz,bb1}, 
we employ light-cone coordinates instead, writing
\be
 \vec{p}^{ }_a =
 p^{ }_{a \parallel} \, \vec{e}^{ }_\vec{k}
 + \vec{p}^{ }_\perp  
 \;, \quad
 \vec{e}^{ }_\vec{k} \; \equiv \; \frac{ \vec{k} }{  k }
 \;. 
\ee
Furthermore we substitute 
$\vec{p}^{ }_a \to -\vec{p}^{ }_a$
in the latter term of \eq\nr{ex5}. 
Then the energies of the particles
of types $a^{ }$ and $b^{ }$ take the forms
\be
 \epsilon^{ }_a \; = \; \sqrt{p_{a\parallel}^2 + p_\perp^2 + m_a^2}
 \;, \quad
 \epsilon^{ }_b \; = \; \sqrt{(k - p^{ }_{a\parallel})^2 + p_\perp^2 + m_b^2}
 \;, \quad
 p^{ }_\perp \; \equiv \; |\vec{p}^{ }_\perp| 
 \;. \la{energies}
\ee
The arguments of the Dirac-$\delta$'s from \eq\nr{ex5} become
\be
 (\omega \mp \epsilon^{ }_a)^2
                 - \epsilon^{2}_{b(\vec{k} - \vec{p}_a)} 
 = 
 M^2 + m_a^2 - m_b^2 \mp 2 \omega \epsilon^{ }_a + 2 k p^{ }_{a\parallel}
 \;, 
 \la{e_constraint} 
\ee
which when put to zero 
establish the relation of $p^{ }_{a\parallel}$
and $\epsilon^{ }_a$ (cf.\ \eq\nr{parallel}).

Writing the integration measure as 
$
 \int \! {\rm d}^3\vec{p}^{ }_a 
 = 
 \int \! {\rm d}^2\vec{p}^{ }_\perp 
 \int_{-\infty}^{\infty} \! {\rm d} p^{ }_{a\parallel}
$, 
the sign of $p^{ }_{a\parallel}$ 
can be dealt with by a special representation of the integrand. 
Inside $\Theta$, we envisage that 
$p^{ }_{a\parallel}$ is solved for by
setting \eq\nr{e_constraint} to zero.
The function $\Theta$ is thereby expressed 
as a function of $\epsilon^{ }_a$ and~$\vec{p}^{ }_\perp$, 
where in turn 
$\epsilon^{ }_a$ is an even function of $p^{ }_{a\parallel}$,
through \eq\nr{energies}.
We can then write
\ba
 \int_{-\infty}^{\infty} \! {\rm d}p^{ }_{a\parallel} \, 
 \delta(\Delta + 2 k p^{ }_{a\parallel} ) \, \phi(p_{a\parallel}^2)
 & = & 
 \int_{0}^{\infty} \! {\rm d}p^{ }_{a\parallel} \, 
 \bigl[ 
 \delta(\Delta + 2 k p^{ }_{a\parallel} ) 
 + 
 \delta(\Delta - 2 k p^{ }_{a\parallel} ) 
 \bigr]
 \, \phi(p_{a\parallel}^2)
 \nn 
 & = & 
 \int_{0}^{\infty} \! {\rm d}p^{ }_{a\parallel} \, 
 4 k p^{ }_{a\parallel} \, 
 \delta(\Delta^2 - 4 k^2 p^{2}_{a\parallel} ) 
 \, \phi(p_{a\parallel}^2)
 \;. 
\ea
Inserting this into \eq\nr{ex5}, and denoting
$
 \int_{\vec{p}_\perp} \equiv \int \! 
 {{\rm d}^2\vec{p}^{ }_\perp} / {(2\pi)^2}
$, we are faced with  
\ba
 & & \hspace*{-1.5cm}
 \Gamma^\rmi{Born}_{1\leftrightarrow 2}
 \; = \; 
 \int_{\vec{p}_\perp}
 \int_0^\infty \! {\rm d} p_{a\parallel} \, p_{a\parallel} \, k
 \biggl\{ 
 \nn 
 & + & 
  \frac{ 
     \sign(\omega - \epsilon^{ }_a) 
     \bigl[1+ n^{ }_{\sigma_a}(\epsilon^{ }_a - \mu^{ }_a)
            + n^{ }_{\sigma_b}(\omega - \epsilon^{ }_a - \mu^{ }_b) \bigr]
       } {\epsilon^{ }_a}
 \nn 
 & \times & 
 \delta\Bigl[ (\MM + m_a^2 - m_b^2 - 2 \omega\epsilon^{ }_a)^2 
 - 4 k^2 p_{a\parallel}^2  
 \Bigr]
 \, 
  {\Theta(\epsilon^{ }_a,\vec{p}^{ }_a,
            \omega - \epsilon^{ }_a,\vec{k}-\vec{p}^{ }_a ) }
 \nn[2mm] 
%%%%%%%%%%%%%
 & - &
 \frac{
     \sign(\omega + \epsilon^{ }_a) 
     \bigl[1 + n^{ }_{\sigma_a}(-\epsilon^{ }_a - \mu^{ }_a)
            + n^{ }_{\sigma_b}(\omega + \epsilon^{ }_a - \mu^{ }_b) \bigr]
      } {\epsilon^{ }_a}
 \nn 
 & \times & 
 \delta\Bigl[ (\MM + m_a^2 - m_b^2 + 2 \omega\epsilon^{ }_a)^2 
 - 4 k^2 p_{a \parallel}^2  
 \Bigr] \, 
 { \Theta(-\epsilon^{ }_a,\vec{p}^{ }_a,
              \omega + \epsilon^{ }_a,\vec{k}-\vec{p}^{ }_a ) }
 \biggr\} 
 \;. \la{1to2_step0}
\ea

Now, by making use of \eq\nr{energies}, we can replace
$(\vec{p}^{ }_\perp,p^{ }_{a\parallel})$ 
as integration variables through 
$(\vec{p}^{ }_\perp,\epsilon^{ }_a)$.
Then, 
we substitute $\epsilon^{ }_a \to -\epsilon^{ }_a$ in the second
structure of \eq\nr{1to2_step0}. The overall minus sign must be 
accounted for, whereby the integrand is weighted by 
$
 \sign\bigl( \epsilon^{ }_a(\omega - \epsilon^{ }_a) \bigr) 
$.
Inside the Dirac-$\delta$, we set
$
 p_{a\parallel}^2 = \epsilon_a^2 - p_\perp^2 - m_a^2
$. 
Finally, we pull out a common factor
$
 {8 k^2 \epsilon^{ }_a(\omega - \epsilon^{ }_a)} / {\omega}
$
from the argument of the Dirac-$\delta$, whereby
the weight function gets multiplied by 
$
  \frac{\omega}{8 k^2 |\epsilon^{ }_a(\omega - \epsilon^{ }_a)|} 
$.
Altogether this yields 
\ba
 & & \hspace*{-1.5cm}
 \Gamma^\rmi{Born}_{1\leftrightarrow 2}
 \; = \; 
 \frac{\omega}{8 k}
 \biggl\{ \int_{-\infty}^{-m_a} + \int_{m_a}^{\infty} \biggr\}
 \frac{{\rm d}\epsilon_a}{\epsilon_a(\omega - \epsilon_a)}
 \nn 
%%%%%%%%%%%%%
 &  \times & 
 \int_{|\vec{p}_\perp| < \sqrt{\epsilon_a^2 - m_a^2 }}
 \, 
     \bigl[1+ n^{ }_{\sigma_a}(\epsilon^{ }_a - \mu^{ }_a)
            + n^{ }_{\sigma_b}(\omega - \epsilon^{ }_a - \mu^{ }_b) \bigr]
 \nn 
 & \times & 
 \delta\biggl[\,  
    \frac{p_\perp^2}{2\epsilon_a} 
  + \frac{p_\perp^2}{2(\omega - \epsilon_a)} 
  + \frac{\omega\MM
    (\epsilon_a - \epsilon_a^{-})
    (\epsilon_a - \epsilon_a^{+})
    }{2 k^2 \epsilon_a (\omega - \epsilon_a)}
 \,\biggr]
 \, \Theta(\P^{ }_a,\K - \P^{ }_a) 
 \;, \la{1to2_step1}
\ea
where we have denoted
\ba
 \epsilon_a^{\pm} & \equiv & 
 \frac{\omega(  \MM + m_a^2 - m_b^2) \pm k \kallen({\MM},m_a^2,m_b^2)}{2\MM}
 \;, \la{epm} 
\ea
with the K\"all\'en function given by
\be
 \kallen^{ } ( M^2,m_a^2,m_b^2 )
  \; \equiv \; 
 \sqrt{     M^4+m_a^4+m_b^4
              - 2 M^2 ( m_a^2  + m_b^2 )
              - 2 m_a^2 m_b^2 
      }
 \;. 
\ee
The longitudinal momentum components, appearing inside $\Theta$, 
satisfy \eq\nr{e_constraint}, {\it viz.} 
\be
  p^{ }_{a\parallel} = 
  \biggl[
    \epsilon^{ }_a + \frac{m_b^2 - m_a^2 - M^2}{2\omega}
  \biggr]\frac{\omega}{k}
  \;, \quad 
  k - p^{ }_{a\parallel} = 
  \biggl[
   \omega - \epsilon^{ }_a 
  + \frac{m_a^2 - m_b^2 - M^2}{2\omega} 
  \biggr]\frac{\omega}{k}
  \;. \la{parallel}
\ee
As guaranteed by the Dirac-$\delta$ in 
\eq\nr{1to2_step1}, $\epsilon^{ }_a$ and $p_\perp^2$ are not independent;
their relation can also be expressed as 
\be
 \epsilon_a^2 - p_{a\parallel}^2 = 
 p_\perp^2 + m_a^2 
 \;, \quad
 (\omega - \epsilon^{ }_a)^2 - (k - p^{ }_{a\parallel})^2 = 
 p_\perp^2 + m_b^2 
 \;. \la{perp}
\ee

Inspecting 
the smallest and largest values of $p_\perp^2$
in \eq\nr{1to2_step1}, 
the Dirac-$\delta$ gets realized
if
\be
 (\epsilon_a - \epsilon_a^{-})
 (\epsilon_a - \epsilon_a^{+}) < 0 
 \;\wedge\;
 \epsilon_a^2 - m_a^2 
 + \frac{M^2}{k^2}(\epsilon_a - \epsilon_a^{-})
                  (\epsilon_a - \epsilon_a^{+}) > 0
 \;. 
\ee 
The latter condition can be completed into a square, and is 
thus always satisfied. This implies that actually no upper bound needs to be 
imposed on $p^{ }_\perp$. The former constraint sets the viable range as
$
 \epsilon_a^{-} < \epsilon^{ }_a < \epsilon_a^{+}
$. 
We also note that 
$ 
 \epsilon^{ }_a > m^{ }_a
$ 
if 
$
 M > m^{ }_a + m^{ }_b
$
or 
$ 
 m^{ }_a > M + m^{ }_b
$, 
and 
$ 
 \epsilon^{ }_a < - m^{ }_a
$
if 
$
 m^{ }_b > M + m^{ }_a
$. 
The domain 
$-m^{ }_a < \epsilon^{ }_a < m^{ }_a$ 
gives no contribution, and does not need to be 
explicitly excluded in \eq\nr{1to2_step1}. 
Making also an effort to write the argument
of the Dirac-$\delta$ in a more transparent form, 
we thus end up with  
\ba
 \Gamma^\rmi{Born}_{1\leftrightarrow 2}
 & = & 
 \frac{\omega}{8 k}
 \int_{-\infty}^{\infty} 
 \frac{{\rm d}\epsilon^{ }_a}{\epsilon^{ }_a(\omega - \epsilon^{ }_a)}
 \,
     \bigl[1+ n^{ }_{\sigma_a}(\epsilon^{ }_a - \mu^{ }_a)
            + n^{ }_{\sigma_b}(\omega - \epsilon^{ }_a - \mu^{ }_b) \bigr]
% \nn 
% & \times & 
 \int_{\vec{p}_\perp} \!\!
 \Theta(\P^{ }_a,\K - \P^{ }_a) % \big|^{ }_{ }
 \nn 
 & \times & 
 \delta \biggl[  
    \frac{ 
       p_\perp^2 + m_a^2 
    +  \frac{ (m_b^2 - m_a^2 - \MM )^2 }{  4 k^2 }  
    }{2\epsilon^{ }_a} 
 +  
    \frac{ 
       p_\perp^2 + m_b^2 
    + \frac{(m_a^2 - m_b^2 - \MM)^2}{4 k^2} 
   }{2(\omega - \epsilon^{ }_a)} 
 -  \frac{\omega\MM}{2 k^2}
 \biggr] 
 \;. \la{1to2_step2} 
\ea
For a polynomial $\Theta$, 
the remaining integrals could be carried out in terms
of polylogarithms, however for us it 
is advantageous to leave them unintegrated. 

%%%%%%%%%%%%%%%%%%%%%%%%%%% SUBSECTION %%%%%%%%%%%%%%%%%%%%%%%%%%%%%%%%%%%
%
\subsection{Examples of matrix elements squared}

As a first example of $\Theta$ in \eq\nr{1to2_step2}, 
consider the production of right-handed neutrinos from 
the symmetric phase of a Standard Model plasma~\cite{bb1}. 
In the 
$1\leftrightarrow 2$ process, the Standard Model particles participating
in this reaction are leptons ($\ell$) and scalars ($\phi$). Being massive,
the right-handed neutrinos can be produced 
with positive or negative helicity ($\tau = \pm$)~\cite{cptheory}. 
The corresponding rate, modulo overall normalization,
can be expressed as
\be
 \Gamma^{\rmi{Born}(\tau)}_{1\leftrightarrow 2}
 = 
 \scat{1\leftrightarrow 2}(\ala,\aS)
 \,\Theta^{\tau}_{ }(\P^{ }_{\ala},\P^{ }_{\cS})
 \;. \la{M_Born_rhn}
\ee
For negative helicity, as is carried by massless Standard Model leptons, 
the rate is suppressed by the right-handed neutrino
mass, {\it viz.}\footnote{%
 To be precise, the ``rate'' we consider here is 
 $ \im [ \bar{u}^{ }_{\vec{k}\tau}
         a^{ }_\rmiii{L} \Pi^\rmiii{R}_{a} \, a^{ }_\rmiii{R} 
     u^{ }_{\vec{k}\tau} ]  
 $, 
 where $ \Pi^\rmiii{R}_{a} $ 
 is the retarded correlator associated with the 
 current 
 $
  \tilde\phi^\dagger \ell^{ }_a
 $, 
 and $a\in\{e,\mu,\tau\}$ 
 is an active lepton flavour. 
 The actual production or equilibration rate is obtained    
 by multiplying this by $h_\nu^2/\omega$, where 
 $h^{ }_\nu$ is a neutrino Yukawa coupling. 
 \la{rhn}
 }  
\be
 \Theta^{-}_{ } = 
 2 (\omega - k )(\epsilon^{ }_{\ala} + p^{ }_{\ala\parallel})
 \;, \la{ThetaM}
\ee
where $p^{ }_{\ala\parallel}$ is given by \eq\nr{parallel}, and 
in the symmetric phase $m^{ }_a = m^{ }_{\ala} = 0$. 

Now, when we make contact with LPM resummation (cf.\ \se\ref{se:LPM}), 
we need to consider $\Theta$ in the UR limit. 
This is defined by assuming that all particle masses are small compared 
with the momenta of the particles
($m_i^2,M^2 \ll \epsilon^2_i, k^2$).
In the UR regime, $\omega \approx k + M^2 / (2k)$, 
and then \eq\nr{parallel} implies that 
$ p^{ }_{a\parallel} \approx \epsilon^{ }_a$. 
Therefore the negative helicity production rate can be estimated as 
$
 \Theta^{-}_{ } 
 \stackrel{\rmiii{UR}}{\approx}
 2 M^2 \epsilon^{ }_\ala / \omega
$, 
where we went back to ``energy-like'' variables in the end. 

For positive helicity, the right-handed neutrinos are in their 
natural state, whereas 
the active leptons experience ``chiral suppression''.
This can be lifted through 
% a vacuum mass or through 
angular momentum transfer, as is manifested by
\be
 \Theta^{+}_{ } 
        = 2 (\omega + k )(\epsilon^{ }_\ala - p^{ }_{\ala\parallel})
        = \frac{2
          (\omega + k ) \, p_\perp^2  }
          { \epsilon^{ }_\ala + p^{ }_{\ala\parallel} }
 \;, \la{ThetaP}
\ee
where we made use of \eq\nr{perp}. 
In the UR regime, this reduces to 
$
 \Theta^{+}_{ } 
 \stackrel{\rmiii{UR}}{\approx}
 2 \omega p_\perp^2 / \epsilon^{ }_\ala
$.

As a second example, we consider photon or dilepton production from 
a QCD plasma~\cite{gelis3,amy1,agmz}. 
Now the rate can be expressed as\hspace*{0.2mm}\footnote{%
 To be precise, the ``rate'' we consider here is 
 $\im [\Pi^{\rmiii{R}\mu\nu}_{ } ] $, where 
 $ \Pi^{\rmiii{R}\mu\nu}_{ } $ is the retarded correlator 
 associated with the vector current $\bar\psi \gamma_{ }^\mu \psi$.
 The actual photon or dilepton production rate 
 is obtained by multiplying this by a kinematic prefactor and 
 by an appropriate power of electromagnetic couplings. 
 \la{dilepton}
 } 
\ba
 \Gamma^{\rmi{Born}({\mu\nu})}_{1\leftrightarrow 2} 
 & = &   
 \scat{1\leftrightarrow 2}(q,\bar{q}) 
 \, \Theta^{\mu\nu}(\P^{ }_q,\P^{ }_{\bar{q}})
 \;, \la{Gamma_munu} \\[2mm]
%%%%%%%% 
 \Theta^{\mu\nu}(\P^{ }_q,\P^{ }_{\bar{q}}) 
 & = & 
% \bigl\{
  2\Nc^{ }
  \, 
  \bigl[
  2 \bigl( \P^\mu_q \P^\nu_{\bar{q}} + \P^\nu_q \P^\mu_{\bar{q}} \bigr)
  - \eta^{\mu\nu}_{ } \, M^2 
  \bigr] 
% \bigr\}
 \;, 
 \la{Theta_munu}
\ea
where $q$ denotes a (possibly massive) quark, 
$\bar{q}$ an antiquark, and 
$\eta^{\mu\nu}_{ } \equiv \diag(\mbox{$+$$-$$-$$-$})$. 
For on-shell photon production, 
we are interested in the transverse projection
\be
 \Gamma^{\rmi{Born}(\rmii{T})}_{1\leftrightarrow 2}
 \; \equiv \; 
 \biggl(\delta^{ }_{ij} - \frac{k^{ }_i k^{ }_j}{k^2} \biggr)
 \Gamma^{\rmi{Born}({ij})}_{1\leftrightarrow 2} 
 \;, \la{GammaT}
\ee
whereas for dileptons the longitudinal components need to be added, 
\be
 \Gamma^{\rmi{Born}(\rmii{L})}_{1\leftrightarrow 2}
 \; \equiv \; 
 \frac{k^{ }_i k^{ }_j}{k^2} \, 
 \Gamma^{\rmi{Born}({ij})}_{1\leftrightarrow 2} 
 - \Gamma^{\rmi{Born}({00})}_{1\leftrightarrow 2}
 \; = \;
 \frac{M^2}{k^2}\,  \Gamma^{\rmi{Born}({00})}_{1\leftrightarrow 2}
 \;. \la{GammaL}
\ee
For the latter representation, we made use of the 
Ward identity 
$ 
 \K^{ }_\mu \Gamma^{\mu\nu}_{1\leftrightarrow 2} = 0
$.
The vector correlator is the sum of the transverse and
longitudinal ones,
\be
 \Gamma^{\rmii{V}}_{ }
 \; \equiv \; 
 \Gamma^{\rmii{T}}_{ } + \Gamma^{\rmii{L}}_{ } 
 = 
 (-\eta^{ }_{\mu\nu}) \Gamma^{({\mu\nu})}_{ }
% \;, \quad  \eta = \diag(\mbox{$+$$-$$-$$-$})
 \;. \la{GammaV} 
\ee

Consider first the longitudinal polarization. 
Employing the latter representation in \eq\nr{GammaL}, 
and inserting \eq\nr{Theta_munu}, the weight function becomes 
\be
 \Theta^\rmii{L}_{ }
 = 
 \frac{ 
 2\Nc^{ } M^2 \bigl( 4 \epsilon^{ }_q \epsilon^{ }_{\bar{q}} - M^2 \bigr)
 }{k^2} 
 \;. \la{ThetaL}
\ee
In the UR regime, this reads 
$
 \Theta^\rmii{L}_{ } 
 \stackrel{\rmiii{UR}}{\approx} 
 8 \Nc^{ }M^2 \epsilon^{ }_q \epsilon^{ }_{\bar{q}} / \omega^2
$, 
rendering a case similar to $\Theta^{-}_{ }$ above. 
 
For the transverse channel, recalling  
$\vec{p}^{ }_{b\perp} = - \vec{p}^{ }_{a\perp}$, 
\eqs\nr{Theta_munu} and \nr{GammaT} yield 
\be
 \Theta^\rmii{T}_{ }
 = 
 4\Nc^{ }
 \bigl(
   - 2 p_\perp^2 + M^2
 \bigr)
 \;. \la{ThetaT}
\ee
This case differs from those considered before, as $p_\perp^2$ appears. 
As dictated by \eq\nr{1to2_step2}, 
at the Born level the magnitude of $p_\perp^2$ is related
to the energies and masses. 
In particular, in the UR regime, $M^2$ and $p_\perp^2$ 
are of the same order. Then we can eliminate
$M^2$ in favour of $p_\perp^2$ through the Dirac-$\delta$ constraint 
in \eq\nr{1to2_step2}, which leads to  
the approximate form that often appears in literature, 
$
 \Theta^\rmii{T}_{ } 
 \stackrel{\rmiii{UR} % ,\rmii{$d=3$}
                      }{\approx}
 4 \Nc^{ }\, p_\perp^2
 [ {\epsilon^{2}_q +  (\omega - \epsilon^{ }_q)^2 } ] / 
                   [ {\epsilon^{ }_q (\omega - \epsilon^{ }_q)} ]
$.

%%%%%%%%%%%%%%%%%%%%%%%%%%% SECTION %%%%%%%%%%%%%%%%%%%%%%%%%%%%%%%%%%%
%
\section{Leading-order LPM resummation}
\la{se:LPM}

%%%%%%%%%%%%%%%%%%%%%%%%%%% SUBSECTION %%%%%%%%%%%%%%%%%%%%%%%%%%%%%%%%
%
\subsection{Known implementations}
\la{ss:lit}

The goal now is to compare \eq\nr{1to2_step2} 
with the framework of leading-order LPM resummation. 
For the benefit of an impatient reader, 
we first reiterate known formulations,
returning in \se\ref{ss:general} to how the matrix elements squared
appearing in them can be derived. 
All the while, it is important to keep in mind that as LPM resummation is 
viable for UR kinematics, there is latitude in how kinematic
variables are chosen beyond this limit. 

Starting with right-handed neutrinos~\cite{bb1}, 
but resolving the helicity channels~\cite{cptheory};
undoing the normalization by $\omega$ that is often
invoked in the literature
(cf.\ footnote~\ref{rhn});
and making a few substitutions 
$k\to\omega$ to render
\eqs\nr{1to2_lpm}, \nr{H} close in appearance to 
\eqs\nr{dilepton_lpm}, \nr{hatH}, 
we can re-express the LPM-resummed result as 
\ba
 \Gamma^{\rmii{LPM}({\tau})}_{1+n\leftrightarrow 2+n } 
%%%%%%%%%%%%%
 & = & 
  \frac{1}{8} 
  \int_{-\infty}^{\infty}   
  \frac{ {\rm d}\epsilon^{ }_\ala }
       {\epsilon^{ }_\ala( \omega - \epsilon^{ }_\ala ) }  
 \bigl[ 1 - \nF^{ }(\epsilon^{ }_\ala - \mu^{ }_\ala )
          + \nB^{ }(\omega - \epsilon^{ }_\ala - \mu^{ }_\aS ) \bigr] 
 \nn[2mm]  
 & \times & 
 \lim_{\vec{y}_{\!\perp}^{ } \to \vec{0}} \mathbbm{P} 
 \biggl\{
  \frac{2 \MM  \epsilon^{ }_\ala\, \delta^{ }_{\tau,-} }{\omega}
  \frac{ \im\, \bigl[g_{ }^{ } (\vec{y}_{\!\perp}^{ } )\bigr] }{\pi}
   \;  +  \;
  \frac{2 \omega\, \delta^{ }_{\tau,+} }{\epsilon^{ }_\ala}
  \frac{ \im\, \bigl[\nabla_\perp\cdot \vec{f}_{ }^{ }
   (\vec{y}_{\!\perp}^{ } )\bigr] }{ \pi }
 \biggr\}
 \;, \hspace*{3mm} \la{1to2_lpm}
\ea
where $\mathbbm{P}$ stands for a principal value,  
and $g_{ }^{ }$ and $\vec{f}_{ }^{ }$ are wave functions satisfying
\ba
 && (\hat{H}^{ }_{ } - i 0^+)\, g^{ }(\vec{y}_{\!\perp}^{ }) \, = \, 
  \delta^{(2)}(\vec{y}_{\!\perp}^{ }) \;, \quad 
 (\hat{H}^{ }_{ } - i 0^+)\, \vec{f}^{ }(\vec{y}_{\!\perp}^{ }) \, = \, 
  -\nabla^{ }_\perp \delta^{(2)}(\vec{y}_{\!\perp}^{ }) 
 \;, \la{Seq} \\[3mm]
 && 
 \hat{H}^{ }_{ } \; \equiv \; 
   \frac{\delta m_{\ell\T}^2 - \nabla_\perp^2}{2\epsilon^{ }_\ala}
 + 
   \frac{m_{\aS\T}^2 
   - \nabla_\perp^2}{2(\omega - \epsilon^{ }_\ala)}
 - \frac{M^2_{ }}{2 \omega}
  \; - \; 
    i \, \sum_{i=1}^{2} 
 g_{\rmii{E}i}^2\, C^{ }_i \, 
 \phi(m^{ }_{\rmii{E}i} y^{ }_\perp)
 \;. \la{H}
\ea
Here 
$
 \delta m_{\ell\T}^{2} \approx (g_1^2 C^{ }_1 + g_2^2 C^{ }_2) T^2 / 4 
 % ( T^2 + \mu^2_\ala / \pi^2 )/16
$ 
is an ``asymptotic'' thermal lepton mass~\cite{weldon}
(quadratic appearances of chemical potentials have been omitted), 
with 
$
 C^{ }_1  \equiv  1/4
$,
$
 C^{ }_2 \equiv  3/4
$; 
$
  m_{\phi\T}^2 \approx  - {\mH^2} / {2} + 
  ( g_1^2 + 3 g_2^2 + 4 h_t^2  + 8 \lambda  
  ){T^2} / {16}
$
is a thermal Higgs mass~\cite{meg}; 
and the thermal width accounts for soft gauge scatterings~\cite{sum1}, 
\ba
%%%%%%%%%%%%%%%%%%%%%%%%
  \phi(\mE y^{ }_\perp) \!\! & \equiv & \!\!
  \int_{\vec{q}_\perp}\!\!
  \bigl( 1 - e^{i \vec{q}^{ }_\perp \cdot \vec{y}_{\!\perp}^{ } }\bigr)
  \biggl( 
   \frac{1}{q_\perp^2} - \frac{1}{q_\perp^2 + \mE^2}
  \biggr)
  = 
  \frac{1}{2\pi}
  \biggl[ \ln\biggl( \frac{ \mE y^{ }_\perp }{2} \biggr)
 + \gammaE + K^{ }_0 \bigl( \mE y^{ }_\perp \bigr) \biggr]
 \;, \nn 
 \la{phi}
\ea
where $K^{ }_0$ is a modified Bessel function.  
The Debye masses 
\be
 m^{2}_\rmii{E1} \; \approx \; 
 \Bigl( \fr{\nS}6 + \frac{5\nG}{9} \Bigr) g_1^2 T^2 
 \;, \quad
 m^{2}_\rmii{E2} \; \approx \; 
 \Bigl( \fr23 + \fr{\nS}6 + \frac{\nG}{3} \Bigr) g_2^2 T^2
 \;, \quad
 \nS \equiv 1 \;, \quad
 \nG \equiv 3 
 \;, 
\ee
and the gauge couplings $g^2_{\rmii{E}i} \approx g_i^2 T$ are
those of the dimensionally reduced effective theory~\cite{sch}. 

% We remark that the structures appearing can be generalized to 
% the Higgs phase, whereby $\gamma$ becomes a ${4\times 4}$ 
% matrix~\cite{broken}.

We note from \eqs\nr{H}, \nr{phi} 
that the parametric magnitude of the thermal width
is $\sim g_2^2 T / \pi$. If 
$\delta m_{\ala\T}^2 / \epsilon^{ }_\ala$, 
$m_{\aS\T}^2 / (\omega - \epsilon^{ }_\ala)$ or 
$M^2 / \omega $ is much larger than this, then the width
can be omitted ($\phi \to 0^+$). 
Then \eq\nr{Seq} can be solved with 
Fourier transformations. 
Recalling the UR limits of $\Theta^{-}_{ }$ and 
$\Theta^{+}_{ }$ from below 
\eqs\nr{ThetaM} and \nr{ThetaP}, respectively, 
we find that in this situation there is a perfect match between 
\eq\nr{1to2_step2} and \eqs\nr{1to2_lpm}, \nr{H}. 

A similar exercise is possible 
for photons and dileptons produced from a massless QCD plasma. 
Following ref.~\cite{agmz}, one viable representation, 
with $i=T,L$, reads
\ba
% && \hspace*{-1.5cm}
 \Gamma^{\rmii{LPM}({i})}_{1+n\leftrightarrow 2+n}
 & \equiv &  
 \Nc
 \int_{-\infty}^{\infty} \! 
 {\rm d}\epsilon^{ }_q  \, 
 \bigl[ 1-\nF^{ }(\epsilon^{ }_q - \mu^{ }_q)
         -\nF^{ }(\omega-\epsilon^{ }_q + \mu^{ }_q) \bigr]
 \nn[2mm] 
%%%%%%%%%%%%%%%%
 & \times &  
 \lim_{\vec{y}_{\!\perp}^{ } \to \vec{0}}  \mathbbm{P} 
 \biggl\{ 
   \frac{M^2 
   \delta^{ }_{i,\rmii{$L$}}
   }{\omega^2}
  \frac{ \im [g(\vec{y}_{\!\perp}^{ } )] }{\pi}
  + 
   \frac{[ \epsilon^2_q + (\omega^{ } - \epsilon^{ }_q)^2 ] 
   \,\delta^{ }_{i,\rmii{$T$}}
    }
        {2 \epsilon^{2}_q (\omega^{ } - \epsilon^{ }_q)^2}  
   \frac{ \im [\nabla^{ }_{\perp}\cdot \vec{f}(\vec{y}_{\!\perp}^{ })] }{\pi} 
 \biggr\} 
 \;,  \hspace*{5mm} \la{dilepton_lpm}  
\ea
where 
$g$ and $\vec{f}$ are Green's functions in the sense of \eq\nr{Seq}. 
Given that the quark and antiquark are degenerate, 
the operator $\hat{H}$ can be simplified into
\be
 \hat{H} = 
   \frac{\omega^{ }
  ( m_\infty^2 - \nabla_\perp^2 )
  }{2\epsilon^{ }_q(\omega^{ }- \epsilon^{ }_q)}
 - \frac{M^2}{2\omega^{ }}
 - i g_\rmii{E3}^2 \CF 
  \, \phi(m^{ }_{\rmii{E3}} y^{ }_\perp)
 \;, \la{hatH}
\ee
where
$
 m_\infty^2
 \equiv \delta m_{q\T}^2
  = \delta m_{\bar{q}\T}^2
  \approx {g_3^2 \CF T^2} / {4}
$ 
is the asymptotic quark thermal mass;
$\CF \equiv (\Nc^2 - 1)/(2\Nc)$;  
and 
$ 
 m^2_{\rmii{E3}} \approx 
 ( {\Nc^{ } }/{ 3 } + { \Nf^{ } }/{6} )
 g_3^2 T^2 
$
as well as 
$g^2_{\rmii{E3}} \approx g_3^2 T$ 
are parameters of the dimensionally reduced theory~\cite{sch}.

Once again, \eqs\nr{dilepton_lpm} and \nr{hatH} agree with 
\eq\nr{1to2_step2} in the limit 
$M^2 / \omega, m_\infty^2/\epsilon^{ }_q \gg g_3^2 T/\pi$, 
if we make use of the UR limits for the two polarization states, 
as given below \eqs\nr{ThetaL} and \nr{ThetaT}, respectively. 

%%%%%%%%%%%%%%%%%%%%%%%%%%% SUBSECTION %%%%%%%%%%%%%%%%%%%%%%%%%%%%%%%%
%
\subsection{Determination of matrix elements squared}
\la{ss:general}

We now return to 
how the matrix elements squared, 
visible on the second rows of \eqs\nr{1to2_lpm} and \nr{dilepton_lpm},
can be derived from the UR limit of the Hard Thermal Loop (HTL) 
effective theory. 
Like in \eq\nr{ex2}, it is sufficient to consider
a $1\to 2$ decay, with the other channels given by
crossings, which are automatically incorporated in the coordinate 
system of \eq\nr{1to2_step2}. 

To obtain these contributions, the resummed
scalar and fermion propagators are needed. The scalar propagator
is simple, as thermal corrections modify the mass but not 
the structure. Then the spectral function (imaginary part
of a retarded propagator) becomes
\be
 \rho^{ }_{\aS}(\P)
 \; \equiv \; 
 - \im \bigl\{ (p^{ }_0 + i 0^+_{ })^2 - \epsilon^{2}_{\aS} \bigr\}^{-1}_{ }
 \; = \; 
 \frac{\pi}{2\epsilon^{ }_{\aS}}
 \bigl[
   \delta(p^{ }_0 - \epsilon^{ }_{\aS}) - 
   \delta(p^{ }_0 + \epsilon^{ }_{\aS})
 \bigr]
 \;, \la{rhoS}
\ee
where we denoted 
$
 \epsilon^2_{\aS} \equiv p^2 + m_{\aS\T}^2
$.
For the fermion propagator, recalling the self-energy 
$\Sigma^{ }_\rmii{HTL}$ within a massless plasma~\cite{weldon,klimov}, 
and taking subsequently the UR limit, we get 
\ba
 \rsl{\rho}^{ }_{\ala}(\P) 
 & \equiv &
 -  
 \im \bigl\{ \bsl{\P} + \bsl{\Sigma}^{ }_\rmii{HTL}
     \bigr\}^{-1}_{p^{ }_0\to p^{ }_0 + i 0^+_{ } }
 %%%%
 \nn 
 & \stackrel{|p^{ }_0|\,>\,p\;\;\;}{=} & 
 \frac{\pi \rsl{n}^{ }_{\!+}}{2}\, 
 \delta\biggl\{ \bigl( p^{ }_0 - p \bigr)( 1 + \tilde{L} )
  - \frac{\delta m_{\ala\T}^2 }{2 p} 
   \biggr\}
 + 
 \frac{\pi \rsl{n}^{ }_{\!-}}{2}\, 
 \delta\biggl\{ \bigl( p^{ }_0 + p \bigr)( 1 - \tilde{L} )
   + \frac{\delta m_{\ala\T}^2 }{2 p} 
 \biggr\}
 %%%%
 \nn 
 & \stackrel{\rmii{UR}}{\approx} & 
 \frac{\pi }{2 \epsilon^{ }_{\ala}}
 \, \Bigl[ 
 \bsl{\overline\P}^{ }_{\!\!\ala+} 
 \,\delta(p^{ }_0 - \epsilon^{ }_{\ala})
 + 
 \bsl{\overline\P}^{ }_{\!\!\ala-} 
 \,\delta(p^{ }_0 + \epsilon^{ }_{\ala})
 \Bigr]
 \;,   \la{rho_ell}
\ea
where 
$
 \tilde L \equiv \delta m_{\ala\T}^2 /({4p^2})
 \ln\{(p^{ }_0 + p)/(p^{ }_0 - p)\}
$, 
$
 \epsilon^{2}_{\ala} \equiv p^2 + \delta m_{\ala\T}^2 
$, 
and 
\be
 n^{ }_\pm \; \equiv \;  (1,\pm\vec{e}^{ }_\vec{p} ) 
 \;, \quad
 \overline{\P}^{ }_{\ala\pm} 
 \; \equiv \; 
 \epsilon^{ }_{\ala}\, n^{ }_{\ala\pm}
 \;, 
\ee
with $\vec{e}^{ }_\vec{p} \equiv \vec{p}/p$ 
standing for a unit vector.\footnote{% 
 The approximation on the last line of \eq\nr{rho_ell} would not 
 be justified if we considered processes very close to threshold, 
 e.g.\ $M \sim m^{ }_{\aS\T}+\delta m^{ }_{\ala\T}$, so that the domain 
 $\epsilon^{ }_{\ala} \sim \delta m^{ }_{\ala\T}$
 could give a substantial contribution.
 } 
Dropping the subscript $\pm$ implies that we consider the ``particle'' mode, 
i.e.\ 
$\overline{\P}^{ }_{\!\!\ala} \equiv \overline{\P}^{ }_{\!\!\ala+}$
and
$n \equiv n^{ }_+$.
For quarks and antiquarks, the spectral functions
have the same form, just with the replacements 
$\delta m_{\ala\T}^2 \to m_\infty^2$, 
$\epsilon^{ }_{\ala} \to \epsilon^{ }_q, \epsilon^{ }_{\bar{q}}$.

Apart from propagators, HTL-resummed vertices are in principle needed
as well. There is no correction to a Yukawa vertex, whereas the coupling
of an electromagnetic current, of momentum $\K$, to a 
thermal quark-antiquark pair, 
takes place via~\cite{htl5,htl6} 
\be
 \Gamma^{\mu}_{ } = 
 \gamma^{\mu}_{ } - 
 \frac{m_\infty^2}{2} \int_{\vec{v}}
 \frac{\mathcal{V}^{\mu}_{ }\rsl{\mathcal{V}} }
 {\mathcal{V}\cdot\P^{ }_q\, \mathcal{V}\cdot\P^{ }_{\bar{q}} }
 \;, \quad
 \mathcal{V} \; \equiv \; (1,\vec{v})
 \;, \la{vertex}
\ee
where the integral goes over the directions of the unit vector $\vec{v}$.

With these rules, we can compute the rates of interest. For right-handed
neutrinos, picking up the particle branches that contribute to the 
$1\to2$ decay
($\equiv \rho^{ }_{\ala +}, \rho^{ }_{\aS +}$), this gives
\ba
% && \hspace*{-1.5cm} 
 \im [ \bar{u}^{ }_{\vec{k}\tau}
         a^{ }_\rmiii{L} \Pi^\rmiii{R}_{a} \, a^{ }_\rmiii{R} 
     u^{ }_{\vec{k}\tau} ]^\rmii{HTL}_\rmi{$1\to2$}
% \nn[2mm]
%%%%%%
 & = & 
 \int_{\P^{ }_{\ala},\P^{ }_{\aS}}
 \hspace*{-1mm} 
 (2\pi)^4 \delta^{(4)}_{ }
 (\K - \P^{ }_{\ala} - \P^{ }_{\aS}) 
  \bigl[ 1 - \nF^{ }(\epsilon^{ }_\ala - \mu^{ }_\ala )
          + \nB^{ }(\epsilon^{ }_\aS - \mu^{ }_\aS ) \bigr]
 \nn[1mm] 
%%%%%%%
 & & \; \times \, 
 4\,
 \tr \{
 \msl{\E}^{\!\tau}_{ } 
 \aL\, \rsl{\rho}^{ }_{\ala +}(\P^{ }_{\ala})\, \aR \,
 \rho^{ }_{\aS +}(\P^{ }_{\aS})
 \}
 \\[2mm] 
%%%%%%%%
 & \equiv & 
 \scat{1\to 2}(\ala,\aS)
 \, \Theta^{\tau}_\rmii{HTL}(\P^{ }_{\ala},\P^{ }_{\aS}) \;, \\[2mm] 
%%%%%%%%
 \Theta^{\tau}_\rmii{HTL}(\P^{ }_{\ala},\P^{ }_{\aS})
 & \stackrel{\rmii{UR}}{\approx} 
 & 4 \E^{\tau}_{ }\cdot {\overline\P}^{ }_{\!\!\ala+}
 \;, \la{scat_htl_rhn}
\ea
where $a^{ }_{\rmii{L(R)}} = (1\mp \gamma^{ }_5)/2$, 
and the helicity projections read
\be
 \E^{+}_{ } \; \equiv \; 
  \frac{\omega + k}{2} \bigl( 1,\vec{e}^{ }_\vec{k} \bigr)
 \;, \quad 
 \E^{-}_{ } \; \equiv \;
  \frac{\omega - k}{2} \bigl( 1,-\vec{e}^{ }_\vec{k} \bigr)
 \;. \la{E_T}
\ee
Eqs.~\nr{scat_htl_rhn} and \nr{E_T} 
give the analogues of \eqs\nr{ThetaM} and \nr{ThetaP},  
respectively, {\it viz.}\
\ba
 \Theta^{-}_\rmii{HTL}
 & \stackrel{\rmii{UR}}{\approx} & \frac{ 2 (\omega - k)\,
     (p^{ }_{\ala} + p^{ }_{\ala\parallel})\,
     \epsilon^{ }_{\ala}
     }{p^{ }_\ala}
 \; \stackrel{\rmii{UR}}{\approx} \; \frac{2M^2 \epsilon^{ }_{\ala}}{\omega}
 \;, \la{ThetaM_htl} \\ 
%%%%%%%%  
 \Theta^{+}_\rmii{HTL}
 & \stackrel{\rmii{UR}}{\approx} & \frac{ 2 (\omega + k) \,
     (p^{ }_{\ala} - p^{ }_{\ala\parallel}) \,
     \epsilon^{ }_{\ala}
     }{p^{ }_\ala}
% \; = \; 
%  \frac{ 2 (\omega + k)
%     \,p^2_{\perp}
%     \, \epsilon^{ }_{\ala}
%     }{
%    (p^{ }_{\ala} + p^{ }_{\ala\parallel})
%    \, p^{ }_\ala
%     }
 \; \stackrel{\rmii{UR}}{\approx} \; 
 \frac{2\omega\, p^2_{\perp}}{\epsilon^{ }_{\ala}}
 \;. \la{ThetaP_htl}
\ea
For the UR limits, we made use of the fact that 
as far as energy-momentum conservation is concerned, 
\eqs\nr{rhoS} and \nr{rho_ell} imply that 
the constraints
take the same form as in vacuum, % {\it viz.}
\be
 \omega = \epsilon^{ }_{\ala} + \epsilon^{ }_{\aS}
 \;, \quad
 \vec{k} = \vec{p}^{ }_{\ala} + \vec{p}^{ }_{\aS}
 \;, \quad
 k\, p^{ }_{\ala\parallel} = 
 \omega\, \epsilon^{ }_{\ala} 
 + \frac{m_{\aS\T}^2 - \delta m_{\ala\T}^2 - M^2}{2} 
 \;, \la{enmom}
\ee
and that therefore 
$
 p^{ }_{\ala} 
 \stackrel{\rmii{UR}}{\approx} 
 p^{ }_{\ala\parallel} 
 \stackrel{\rmii{UR}}{\approx}
 \epsilon^{ }_{\ala}
$.

For the electromagnetic current, mediated by the vertex from 
\eq\nr{vertex}, we similarly get
\ba
% && \hspace*{-1.5cm} 
 \im [\Pi^{\rmiii{R}\mu\nu}_{ } ]^\rmii{HTL}_\rmi{$1\to2$}
% \nn[2mm]
%%%%%%
 & = & 
 \int_{\P^{ }_{q},\P^{ }_{\bar{q}}}
 \hspace*{-1mm} 
 (2\pi)^4 \delta^{(4)}_{ }
 (\K - \P^{ }_{q} - \P^{ }_{\bar{q}}) 
  \bigl[ 1 - \nF^{ }(\epsilon^{ }_q - \mu^{ }_q )
          - \nF^{ }(\epsilon^{ }_{\bar{q}} + \mu^{ }_q ) \bigr]
 \nn[1mm] 
%%%%%%%
 & & \; \times \, 
 2\Nc^{ }\,
 \tr \{
 \Gamma^{\mu}_{ }
 \rsl{\rho}^{ }_{q +}(\P^{ }_{q})\, 
 \Gamma^{\nu}_{ } \,
 \rsl{\rho}^{ }_{\bar{q} +}(\P^{ }_{\bar{q}})
 \}
 \\[2mm] 
%%%%%%%%
 & \equiv & 
 \scat{1\to 2}(q,\bar{q})
 \, \Theta^{\mu\nu}_\rmii{HTL}(\P^{ }_q,\P^{ }_{\bar{q}})
 \;, 
 \\[2mm] 
%%%%%%%%
 \Theta^{\mu\nu}_\rmii{HTL}(\P^{ }_q,\P^{ }_{\bar{q}}) 
 & \stackrel{\rmii{UR}}{\approx} & 
 \frac{\Nc^{ }}{4}
 \Bigl[
  \tr\bigl( \Gamma^{\mu}_{ } \bsl{\overline{\P}}^{ }_{\!\!q} \bigr) 
  \tr\bigl( \Gamma^{\nu}_{ } \bsl{\overline{\P}}^{ }_{\!\!\bar{q}} \bigr) 
 + 
  \tr\bigl( \Gamma^{\mu}_{ } \bsl{\overline{\P}}^{ }_{\!\!\bar{q}} \bigr) 
  \tr\bigl( \Gamma^{\nu}_{ } \bsl{\overline{\P}}^{ }_{\!\!q} \bigr) 
 - 
  4 {\overline{\P}}^{ }_{\!\!q} \cdot {\overline{\P}}^{ }_{\!\!\bar{q}}
  \, 
  \tr\bigl( \Gamma^{\mu}_{ } \Gamma^{\nu}_{ } \bigr) 
 \Bigr]
 \;. \la{Theta_munu_htl} \nn
\ea
The influence of the 
vertex correction 
is tedious to work out. 
It can be shown, however, that the Ward identity 
$\K^{ }_\mu \Theta^{\mu\nu}_\rmii{HTL} = 0$ is satisfied within
the UR approximation, and that for $\Theta^{00}_\rmii{HTL}$ and 
$\Theta^\rmii{T}_\rmii{HTL}$ the vertex corrections can be 
omitted~\cite{agz}, i.e.\ 
we can replace $\Gamma^{\mu}_{ }\to \gamma^{\mu}_{ }$. 
Thanks to the Ward identity we can use 
$
 \Theta_{\rmii{HTL}}^{\rmii{L}}
 = 
 {M^2 \Theta_{\rmii{HTL}}^{00}} / {k^2} 
$.
Then, employing energy-momentum 
conservation like in \eq\nr{enmom}, which implies
$
 \vec{p}^{ }_q \cdot \vec{p}^{ }_{\bar{q}} = 
 \epsilon^{ }_q \epsilon^{ }_{\bar{q}} + m_\infty^2 - {M^2}/{2}
$,
\eq\nr{Theta_munu_htl} leads to 
\ba
 \Theta_{\rmii{HTL}}^{\rmii{L}}
 & \stackrel{\rmii{UR}}{\approx} &  
%%%%
  \frac{ 4 \Nc^{ } M^2 \epsilon^{ }_q \epsilon^{ }_{\bar{q}} 
 \bigl( 
   2 - n^{ }_q \cdot n^{ }_{\bar{q}} 
 \bigr) }{k^2}
 \\ 
%%%%
 & = &  
 \frac{4 \Nc^{ } M^2 \epsilon^{ }_q \epsilon^{ }_{\bar{q}}
  \bigl( 
    p^{ }_q p^{ }_{\bar{q}} + 
    \vec{p}^{ }_q \cdot \vec{p}^{ }_{\bar{q}}
  \bigr)
  }
  {p^{ }_q p^{ }_{\bar{q}} k^2 }
% \nn 
%%%%
% & = & 
% \frac{2 \Nc^{ } M^2 \epsilon^{ }_q \epsilon^{ }_{\bar{q}}
% \bigl[
%  2\bigl(  
%   p^{ }_q p^{ }_{\bar{q}} + 
%  \epsilon^{ }_q \epsilon^{ }_{\bar{q}} + m_\infty^2 
%   \bigr) - M^2 
% \bigr]
% }{p^{ }_q p^{ }_{\bar{q}} k^2}
% \;, \\
%%%%%%%%%%%%%
 \; \stackrel{\rmii{UR}}{\approx} \;
 \frac{8 \Nc^{ }M^2 \epsilon^{ }_q \epsilon^{ }_{\bar{q}}}{\omega^2}
 \;, \la{ThetaL_htl_ur} \\
%%%%%%%%%%%%%
 \Theta_{\rmii{HTL}}^{\rmii{T}}
 & \stackrel{\rmii{UR}}{\approx} &  
%%%%
  8 \Nc^{ } \epsilon^{ }_q \epsilon^{ }_{\bar{q}} 
 \bigl( 
    n^{i}_{q\perp} n^i_{\bar{q}\perp} + n^{ }_q \cdot n^{ }_{\bar{q}} 
 \bigr) 
 \\[2mm] 
%%%%
 & = & 
 \frac{8 \Nc^{ } \epsilon^{ }_q \epsilon^{ }_{\bar{q}}
 \bigl(
    p^{ }_q p^{ }_{\bar{q}} 
  - p^{ }_{q\parallel} p^{ }_{\bar{q}\parallel} 
 \bigr)
 }{p^{ }_q p^{ }_{\bar{q}} }
%%%%%%%%%%%%%
 \; \stackrel{\rmii{UR}}{\approx} \;
 \frac{4 \Nc^{ }\, p_\perp^2 (\epsilon_q^2 + \epsilon_{\bar{q}}^2)}
 {\epsilon^{ }_q\epsilon^{ }_{\bar{q}}}
 \;. \la{ThetaT_htl_ur}
\ea
The last forms in \eqs\nr{ThetaL_htl_ur} and \nr{ThetaT_htl_ur}
coincide with those in \eq\nr{dilepton_lpm}, 
after multiplying with the factor 
$1 / (8 \epsilon^{ }_q \epsilon^{ }_{\bar{q}})$
from the integration measure.

%%%%%%%%%%%%%%%%%%%%%%%%%%% SECTION %%%%%%%%%%%%%%%%%%%%%%%%%%%%%%%%%%%
%
\section{Interpolation}
\la{se:interpolation}

%%%%%%%%%%%%%%%%%%%%%%%%%%%%%%%%% FIGURE %%%%%%%%%%%%%%%%%%%%%%%%%%%%%%%%%
\begin{figure}[t]

\hspace*{-0.1cm}
\centerline{%
 \epsfysize=7.5cm\epsfbox{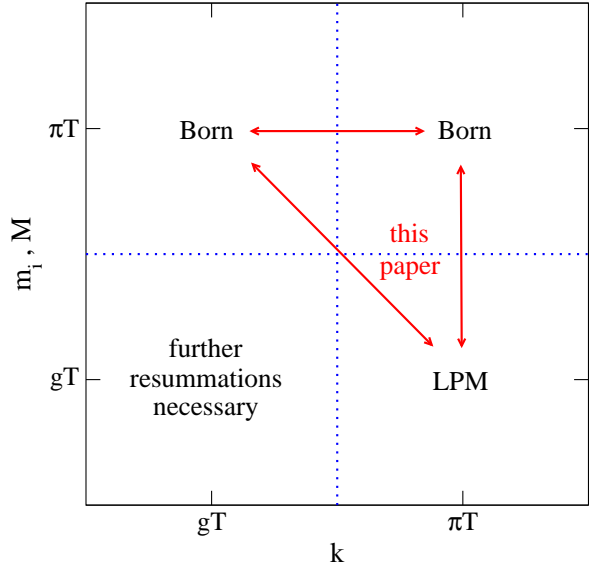}%
}

\caption[a]{\small
 An overview of various momentum and mass domains, 
 with $g \ll \pi$ denoting a generic coupling. 
 The interpolation introduced in
 \se\ref{se:interpolation} can be used in 
 three of the four quadrants shown, but does not apply when all masses
 and momenta are very small (blindly extrapolating the interpolant 
 into that quadrant,
 the result could even become negative, which is unphysical). 
}

\la{fig:overview}
\end{figure}
%%%%%%%%%%%%%%%%%%%%%%%%%%%%%%%%%%%%%%%%%%%%%%%%%%%%%%%%%%%%%%%%%%%%%%%%%%%

The results of \ses\ref{se:1to2} and \ref{se:LPM} are parametrically
correct in different mass and momentum domains, as illustrated in 
\fig\ref{fig:overview}. The purpose of the present section is to
suggest a smooth interpolation between these domains, 
making the results more broadly applicable. 

The key idea is to generalize the Hamiltonians in 
\eqs\nr{H} and \nr{hatH} so that they apply 
beyond the UR limit.\footnote{%
 We note in passing that in principle 
 there could be several ($\equiv n$) non-degenerate $1\to 2$ 
 channels for the non-equilibrium particle to decay into.
 Then $\hat{H}^{ }_\rmii{ }$ would
 become an $n\times n$-matrix, with kinetic terms appearing  
 on the diagonal, whereas the width 
 matrix takes a non-diagonal 
 appearance in this basis~\cite{broken}.
 } 
To this aim we keep 
the kinematic and mass structures from inside the Dirac-$\delta$ in 
\eq\nr{1to2_step2}, and then just replace $p_\perp^2 \to -\nabla_\perp^2$
and add the thermal width, {\it viz.}
\ba
 \hat{H}^{ }_\rmi{inter} & \equiv &  
     \frac{ 
       m_{a\T}^2 - \nabla_\perp^2 
    +  \frac{ (m_{b}^2 - m_{a}^2 - \MM )^2 }{  4 k^2 }  
    }{2\epsilon^{ }_a} 
 +  
    \frac{ 
       m_{b\T}^2 - \nabla_\perp^2
    + \frac{(m_{a}^2 - m_{b}^2 - \MM)^2}{4 k^2} 
   }{2(\omega - \epsilon^{ }_a)} 
 -  \frac{\omega\MM}{2 k^2}
%%%%%%%%%
 \nn 
 & - & i \sum_i g^2_{\rmii{E}i}\, C^{ }_i\, 
   \phi(m^{ }_{\rmii{E}i}\vec{y}_{\!\perp}^{ } )
 \;. \la{H_inter}
\ea
Here $m_{a\T}^2 \equiv m_a^2 + \delta m_{a\T}^2$ and
$m_{b\T}^2 \equiv m_b^2 + \delta m_{b\T}^2$ include thermal corrections, 
such that at high temperatures they go over to asymptotic thermal masses.
The sum over $i$ goes over the gauge representations involved. 
Employing 
$m_{a\T}^2$ and $m_{b\T}^2$ also in the  
     terms      $ (m_{b}^2 - m_{a}^2 - \MM)^2/k^2 $ and 
                $ (m_{a}^2 - m_{b}^2 - \MM)^2/k^2 $
would be formally a higher-order effect in the UR regime, 
nevertheless this procedure can be given a justification
only under specific circumstances 
(cf.\ \se\ref{se:OPE}), 
so for the moment we do not adopt such a recipe.  

It is more subtle to generalize the second lines of 
\eqs\nr{1to2_lpm} and \nr{dilepton_lpm}
beyond the UR limit. After longitudinal momenta are 
eliminated through \eq\nr{parallel}, the matrix elements squared in 
\eqs\nr{ThetaM}, \nr{ThetaP}, \nr{ThetaL}, \nr{ThetaT}
are functions of energy variables, $p_\perp^2$, vacuum masses, 
and the variables $\omega,k,\MM$ that characterize the 
non-equilibrium particle. However, these are not independent
of each other within the Born limit, since
the Dirac-$\delta$ constraint in \eq\nr{1to2_step2}
permits to eliminate $M^2$ in favour of $p_\perp^2$, 
or vice versa. The problem is that once thermal masses
are introduced in \eq\nr{H_inter}, previously equivalent
representations are no longer so, and in fact even their IR 
sensitivities differ. Yet the correct IR structure is unique
and needs to be present 
(cf.\ \ses\ref{ss:general} and \ref{se:subtraction}). 

If the LPM expressions are known
(cf.\ \eqs\nr{1to2_lpm} and \nr{dilepton_lpm}), 
it is possible to make use of the freedom  
in order to choose a representation which 
goes over to the correct LPM one in the UR limit. Concretely, 
noting that terms without $p_\perp^2$ correspond to 
$\lim_{\vec{y}_{\!\perp}^{ } \to \vec{0}} \im [g(\vec{y}_{\!\perp}^{ })]/\pi$
and $p_\perp^2$ to 
$\lim_{\vec{y}_{\!\perp}^{ } \to \vec{0}} 
\im [\nabla^{ }_\perp\cdot \vec{f}(\vec{y}_{\!\perp}^{ })]/\pi$,  
and indicating this transition as 
\be
 \Theta(\P^{ }_a, \K - \P^{ }_a) \; \equiv \; 
 \Theta(\epsilon^{ }_a,p_\perp^2) \; \to \; 
 \widetilde\Theta(\epsilon^{ }_a,\vec{y}_{\!\perp}^{2} )
 \;, \la{transition}
\ee
we can identify\footnote{% 
 For 
 $
  \widetilde\Theta^{-}_{ }
 $, 
 we start from \eq\nr{ThetaM} 
 and insert $p^{ }_{\ala\parallel}$ from \eq\nr{parallel}, 
 with $m^{ }_a \to m^{ }_\ala = 0$ and 
 $m^{ }_b\to m^{ }_{\aS\T}$. 
 For 
 $
  \widetilde\Theta^{+}_{ }
 $, 
 we start from the first form in \eq\nr{ThetaP}, 
 take $p^{ }_{\ala\parallel}$ from \eq\nr{parallel}, 
 pull apart 
 $
  2M^2(\omega - \epsilon^{ }_\ala)/k
 $, 
 and insert there $M^2$ from \eq\nr{1to2_step2}.
 For 
 $
   \widetilde\Theta^\rmii{L}_{ }
 $, 
 the form comes directly from 
 \eq\nr{ThetaL}.
 For 
 $
   \widetilde\Theta^\rmii{T}_{ }
 $, 
 we start from \eq\nr{ThetaT}, 
 and insert $M^2$ from \eq\nr{1to2_step2}, 
 {\it viz.} 
 $
  M^2 = 
  [k^2(p_\perp^2 + m_q^2) + M^4/4]/ 
  [\epsilon^{ }_q(\omega - \epsilon^{ }_q)]
 $. 
 } 
\ba
 \widetilde\Theta^{-}_{ }
    \!\!\!& = &\!\!  
 \frac{ (\omega - k) 
 \bigl[\, 
   2 (\omega + k ) \epsilon^{ }_{\ala}  
 + m_{\aS\T}^2 % - \delta m^2_{\ala\T}
 - M^2 
 \,\bigr]  }{k}
% \biggl\{
   \, 
   \frac{ \im [g(\vec{y}_{\!\perp}^{ } )] }{\pi}
   \,
% \biggr\}
 \;, \la{ThetaM_new} \\[2mm] 
%%%%%%%
 \widetilde\Theta^{+}_{ } 
        \!\!\!& = &\!\! 
          \frac{
          ( m_{\aS\T}^2 - M^2 )
           \bigl[\, 
            2 ( \omega - k ) \epsilon^{ }_{\ala}
           + 
            m_{\aS\T}^2 - M^2
           \,\bigr]
           }{2 k \epsilon^{ }_{\ala}}
%          \biggl\{ 
          \,
          \frac{ \im [g(\vec{y}_{\!\perp}^{ } )] }{\pi}
          \,
%          \biggr\}
          + 
         \frac{2 k}{ \epsilon^{ }_{\ala} }
%          \biggl\{
  \,
  \frac{ \im [\nabla^{ }_{\perp}\cdot \vec{f}(\vec{y}_{\!\perp}^{ } )] }{\pi} 
  \,
%          \biggr\}
 \;, \hspace{8mm} \la{ThetaP_new} \\[2mm] 
%%%%%%%%
 \widetilde\Theta^\rmii{L}_{ }
           \!\!& = &\!\! 
 \frac{ 
 2\Nc^{ } M^2 
 \bigl[ 4 \epsilon^{ }_q \bigl( \omega - \epsilon^{ }_{q} \bigr) - M^2 \bigr]
 }{k^2} 
% \biggl\{
 \, 
    \frac{ \im [g(\vec{y}_{\!\perp}^{ } )] }{\pi}
 \,
% \biggr\} 
 \;, \la{ThetaL_new} \\[2mm]
%%%%%%%%%%
 \widetilde\Theta^\rmii{T}_{ }
         \!\!\!& = &\!\! 
 4\Nc^{ }
 \biggl\{ 
    \frac{ 
     4 k^2 m_q^2 
    + M^4  
   }{4 \epsilon^{ }_q ( \omega - \epsilon^{ }_q ) } 
  \, \frac{ \im [g(\vec{y}_{\!\perp}^{ } )] }{\pi}
 +  
    \frac{ 
     \epsilon^{2}_q +  ( \omega - \epsilon^{ }_q )^2 - \MM     
   }{\epsilon^{ }_q ( \omega - \epsilon^{ }_q ) } 
  \,
   \frac{ \im [\nabla^{ }_{\perp}\cdot \vec{f}(\vec{y}_{\!\perp}^{ } )] }{\pi} 
 \biggr\} 
 \;. \la{ThetaT_new} \hspace*{6mm}
\ea

A few important remarks are in order. 
First, we have replaced the Higgs vacuum mass
by a Higgs thermal mass in \eqs\nr{ThetaM_new} and \nr{ThetaP_new}, 
and do the same in the Born-level expressions; 
otherwise a symmetric phase computation would make no sense.
Second, the denominators 
in \eqs\nr{ThetaP_new} and \nr{ThetaT_new}
can have zeros. 
With the Born-level integration range incorporating finite masses, 
they would be avoided, 
but with a finite width, it is non-trivial to verify
that there is no spectral weight in this domain. 
To be prudent, 
the integrations over~$\epsilon^{ }_{\ala}$ or~$\epsilon^{ }_q$
may be implemented as principal values.  
Third, the representations are not unique, as they could be 
modified by terms that are of higher order in the UR regime. 
It turns out there are consistency conditions between
\eq\nr{H_inter} and \eqs\nr{ThetaM_new}--\nr{ThetaT_new} that need to 
be satisfied; we return to 
this in \se\ref{se:OPE}. The specific forms
in \eqs\nr{H_inter} and 
\nr{ThetaM_new}--\nr{ThetaT_new} have been chosen in order to 
guarantee a straightforward matching in \se\ref{se:subtraction}, 
but we have also successfully
tested another implementation,\footnote{%
 For
 $
  \widetilde\Theta^{-}_{ }
 $,
 we start from the first form of \eq\nr{ThetaM_htl}
 and insert $p^{ }_{\ala\parallel}$ from \eq\nr{enmom}.
 For
 $
  \widetilde\Theta^{+}_{ }
 $,
 we start from the first form in \eq\nr{ThetaP_htl},
 take $p^{ }_{\ala\parallel}$ from \eq\nr{enmom},
 pull apart
 $
  2M^2(\omega - \epsilon^{ }_\ala)\,\epsilon^{ }_\ala/(k\, p^{ }_{\ala})
 $,
 and insert there $M^2$ from \eq\nr{1to2_step2},
 with $m^{ }_a \to \delta m^{ }_{\ala\T}$ and
 $m^{ }_b \to m^{ }_{\aS\T}$. 
 In both  
 $
  \widetilde\Theta^{-}_{ }
 $ 
 and  
 $
  \widetilde\Theta^{+}_{ }
 $, we then set 
 $
   p^{ }_{\ala} \equiv \sign(\epsilon^{ }_\ala)
   \sqrt{\epsilon^{2}_\ala - \delta m_{\ala\T}^2}
 $
 and restrict the integration range
 in \eq\nr{1to2_inter} to 
 $ 
 | \epsilon^{  }_\ala | \ge | \delta m_{\ala\T}^{ } |
 $.
 We recall that for $\epsilon^{ }_\ala\sim \delta m^{ }_{\ala\T}$, 
 none of the approaches of this paper
 represents a correct description of soft physics, 
 however this region gives a subleading
 contribution to \eq\nr{1to2_inter}. 
% thus, different prescriptions do provide 
% an equivalent LO determination of the latter. 
 An NLO computation in the UR regime would require the subtraction
 of the $\epsilon^{ }_\ala\sim \delta m^{ }_{\ala\T}$ region in
 \eq\nr{1to2_inter}, followed by its full-fledged HTL inclusion, 
 as discussed in ref.~\cite{nlo_photon}
 for the electromagnetic case.
 } 
which might instead facilitate a matching 
needed for an NLO computation in the UR regime.

Once the matrix elements / splitting functions have been fixed, 
the interpolation can be expressed as 
\be
 \Gamma^\rmi{inter}_{1+n\leftrightarrow 2+n}
  =  
 \frac{\omega}{8 k}
 \int_{-\infty}^{\infty} 
 \! {\rm d}\epsilon^{ }_a \, 
 \,
     \bigl[1+ n^{ }_{\sigma_a}(\epsilon^{ }_a - \mu^{ }_a)
            + n^{ }_{\sigma_b}(\omega - \epsilon^{ }_a - \mu^{ }_b) \bigr]
 \lim_{\vec{y}_{\!\perp}^{ } \to\vec{0}}
 \mathbbm{P}\biggl\{ 
 \frac{
 \widetilde\Theta(\epsilon^{ }_a,\vec{y}_{\!\perp}^{2} ) 
 }{\epsilon^{ }_a(\omega - \epsilon^{ }_a)}
 \biggr\} 
 \;, \la{1to2_inter} 
\ee
with 
the wave functions solved from \eq\nr{Seq}
and the Hamiltonian taken from \eq\nr{H_inter}. 
Such an interpolation on the integrand level
smoothly connects the LPM-resummed and Born regimes. 
We stress, however, that the result is reliable only
in the Born ($m^{ }_i,M \gg gT$) and UR regimes
($m^{ }_i,M \ll k \sim \pi T $), whereas it should not 
be used for $k, m^{ }_i, M \ll \pi T$ (cf.\ \fig\ref{fig:overview}).

%%%%%%%%%%%%%%%%%%%%%%%%%%% SECTION %%%%%%%%%%%%%%%%%%%%%%%%%%%%%%%%%%%
%
\section{Crosscheck of UV asymptotics}
\la{se:OPE}

According to \fig\ref{fig:overview}, 
the results of the interpolation should go over to the
Born ones at large masses. A special case is when the virtuality of the
probe particle is made large, $M \gg m^{ }_a,m^{ }_b,\pi T$. The structure
of thermal spectral functions is well 
understood in this limit~\cite{ope}, and merits a discussion. 

We start by recalling the vacuum values of the interaction rates 
at the Born level:
\ba
 \Gamma^{\rmi{Born}(-)}_{1\leftrightarrow 2} \bigr|^{ }_{T=0}
 \; = \;
 \Gamma^{\rmi{Born}(+)}_{1\leftrightarrow 2} \bigr|^{ }_{T=0}
 & = & 
 \frac{(\MM + m_{\ala}^2 - m_{\aS}^2)
 \kallen(\MM,m^2_{\ala},m^2_{\aS})}{16\pi\MM}
%%%%%%
 \nn 
 & \stackrel{M \gg m^{ }_{\aS},\;m^{ }_{\ala}=0}{=} & 
 \frac{1}{16\pi}
 \biggl[
   \MM - 2 m^2_{\aS} + \rmO\biggl(\frac{m^4_{\aS}}{\MM}\biggr) 
 \biggr]
 \;, \la{rhn_asy} \\[2mm]
%%%%%%
 \Gamma^{\rmi{Born}(\rmii{T})}_{1\leftrightarrow 2} \bigr|^{ }_{T=0}
 \; = \;
 2\,\Gamma^{\rmi{Born}(\rmii{L})}_{1\leftrightarrow 2} \bigr|^{ }_{T=0}
 & = & 
 \frac{\Nc^{ }(\MM + 2 m_{q}^2)
 \kallen(\MM,m^2_{q},m^2_{q})}{6\pi\MM}
%%%%%%
 \nn 
 & \stackrel{M \gg m^{ }_{q}}{=} & 
 \frac{\Nc^{ }}{6\pi}
 \biggl[
   \MM + \rmO\biggl(\frac{m^4_{q}}{\MM}\biggr) 
 \biggr]
 \;. \la{dilepton_asy}
\ea
Now, if we include thermal mass corrections in the scalar mass,
i.e.\ $m^2_{\aS}\to m^2_{\aS\T} = m^2_{\aS} + \delta m^2_{\aS\T}$, 
then \eq\nr{rhn_asy} suggests the presence of a 
correction of $\rmO(g^2T^2)$. To see the correct behaviour in this limit,
requires a full NLO computation~\cite{salvio,nonrel,biondini}. 
It turns out that 
there is a contribution proportional 
to $\rmO(\lambda T^2 )$, 
however it is not equivalent to $\delta m^2_{\aS\T}$.
In the dilepton case, thermal corrections are of
$\rmO(g^2 T^4/\MM)$~\cite{ope,harvey}, and  
again not related to $\delta m^4_{q\T}$, even if the 
same power of $T$ would be obtained from \eq\nr{dilepton_asy}.

In any case, 
introducing thermal mass corrections in \eq\nr{H_inter}, 
changes the asymptotics of 
$
  \Gamma^\rmi{inter}_{1+n\leftrightarrow 2+n}
$. 
Generically, the changes are 
proportional to $\delta m^2_{a\T}$, $\delta m^2_{b\T}$,  
and such ``offsets''
are also visible in the numerical results of \se\ref{se:numerics}.
As mentioned, these offsets do {\em not} agree with what 
a full NLO computation would yield, but  
there is no reason to worry, 
for the offsets are subleading
corrections in the regime $M \gg gT$. In fact, they could be dealt with 
in connection with a matching, 
as described in \se\ref{se:subtraction}. 

It turns out, 
however, that the precise way in which thermal mass corrections
are introduced in \eq\nr{H_inter}, is very important. 
Without sufficient care, the UV asymptotics would 
display not only offsets, 
but even terms that grow with~$M$, as we now demonstrate.

Considering the asymptotic regime, $M \gg \pi T \gg g^2 T/\pi$,  the width
can be omitted in \eq\nr{H_inter}. Physically, this means that we 
consider the $n=0$ version of \eq\nr{1to2_inter}. Keeping 
separate handles for various appearances of masses, and undoing
the transformation in \eq\nr{transition}, the starting point can be 
written in a form similar to \eq\nr{1to2_step2}, {\it viz.} 
\ba
 \Gamma^\rmi{inter}_{1\leftrightarrow 2}
 \!\! & \equiv & \!\! 
 \int_{-\infty}^{+\infty} 
 \! {\rm d}\epsilon^{ }_a \, 
 \frac{ 
 \sign({\epsilon^{ }_a(\omega - \epsilon^{ }_a)})
 }{16\pi k}
 \,
     \bigl[1+ n^{ }_{\sigma_a}(\epsilon^{ }_a - \mu^{ }_a)
            + n^{ }_{\sigma_b}(\omega - \epsilon^{ }_a - \mu^{ }_b) \bigr]
% \nn 
% & \times & 
 \int_0^\infty \! {\rm d}p_\perp^2 \, 
 \Theta(\epsilon^{ }_a,p_\perp^2)
 \hspace*{3mm}
%%%%%%%%
 \nn 
 \!\! & \times & \!\!
 \delta \biggl\{   
    p_\perp^2 
   - 
   \frac{2\epsilon^{ }_a(\omega - \epsilon^{ }_a)}{\omega}
   \biggl[ 
    \frac{\omega\MM}{2k^2}
  - 
    \frac{ 
       m_{a\T}^2 
    +  \frac{ (m_b^2 - m_a^2 - \MM )^2 }{  4 k^2 }  
    }{2\epsilon^{ }_a} 
  -   
    \frac{ 
        m_{b\T}^2 
    + \frac{(m_a^2 - m_b^2 - \MM)^2}{4 k^2} 
   }{2(\omega - \epsilon^{ }_a)} 
 \biggr] 
 \biggr\} 
 \;. \la{1to2_subtr0}
%%%%%%
\ea
Solving for the integration boundaries by requiring that the 
argument of the Dirac-$\delta$ is infinitesimally negative at 
$p_\perp^2 \to 0$, \eq\nr{epm} gets generalized into
\ba
 \epsilon_{a\T}^{\pm} & \equiv & 
 \frac{\omega(  \MM + m_{a\T}^2 - m_{b\T}^2 )
       + \frac{ M^2 \Delta^{ }_\T }{ \omega } 
       \pm k \sqrt {\lambda({\MM},m_{a\T}^2,m_{b\T}^2)
                   - (\frac{ M \Delta^{ }_\T }{ \omega })^2_{ } }
       }{2\MM}
 \;, \la{epmT} \\
 \Delta^{ }_\T & \equiv & 
 m_a^2 - m_{a\T}^2 + m_{b\T}^2 - m_b^2
 \;. \la{Delta}
\ea
For a fixed $k$ and masses, but taking $M$ large, this can be simplified into
\be
 \epsilon_{a\T}^{\pm} \; \equiv \; \bar{\epsilon}_{a\T}^{ } \pm 
                        \Delta \epsilon_{a\T}^{ }
 \;, \quad
 \bar{\epsilon}_{a\T}^{ } \stackrel{M \gg m^{ }_i,k}{\approx}
 \frac{M}{2} + \frac{m_a^2 - m_b^2 + k^2/2}{2M}
 \;, \quad
 \Delta \epsilon_{a\T}^{ } \stackrel{M \gg m^{ }_i,k}{\approx}
 \frac{k}{2}
 \;. \la{boundaries}
\ee
The important observation is that the average energy is determined by 
the masses $m_a^2$, $m_b^2$, appearing in the ``subleading'' mass
corrections in \eq\nr{1to2_subtr0}. The reason is that at $M\gg k$, 
these terms are enhanced by $M^2/k^2$ compared with the contribution
from $m_{a\T}^2$, $m_{b\T}^2$. 

Inserting now the matrix element from \eq\nr{ThetaM_new},  
with $m_{\aS\T}^2 \to m_b^2$ and re-introducing $m_a^2$ as
it appears in \eq\nr{parallel}, let us denote
\be
 \epsilon^{ }_0 
  \; \equiv \; 
 \frac{M^2 + m_{a}^2 - m_{b}^2}{2(\omega + k)}
 \; \stackrel{M \gg m^{ }_i,k}{\approx} \; 
 \frac{M-k}{2} 
 + 
 \frac{m_{a}^2 - m_{b}^2 + k^2/2}{2M}
 \;. \la{eps0}
\ee
Then we are faced with 
\be
 \Gamma^{\rmi{inter}(-)}_{1\leftrightarrow 2} 
  \stackrel{M \gg m^{ }_i,k,\pi T}{\approx} 
 \int_{\epsilon_{a\T}^-}^{\epsilon_{a\T}^+} 
 \! {\rm d}\epsilon^{ }_a \, 
 \frac{M^2 (\epsilon^{ }_a - \epsilon^{ }_0)  }{8\pi k^2} 
%%% 
 \; = \; 
 \frac{M^2 ( \epsilon^{+}_{a\T} - \epsilon^{-}_{a\T} ) }{8\pi k^2}
 \biggl( \frac{ \epsilon^{+}_{a\T} + \epsilon^{-}_{a\T} }{2}
      - \epsilon^{ }_0 \biggr)
 \;. 
%%%
% \nn 
% & \stackrel{M \gg m^{ }_i,k}{\approx} & 
% \frac{1}{16\pi} 
% \biggl[
%    M^2 + \frac{M \Delta^{ }_\T }{k} 
%   + \rmO(1) 
% \biggr]
% \;.
\ee
Inserting \eqs\nr{boundaries} and \nr{eps0}, 
the leading term reproduces \eq\nr{rhn_asy}. 
However, if the masses in the $\rmO(1/M)$-parts did not match, 
the subleading correction would be of $\rmO(M)$.

To summarize, in order to guarantee the presence of vacuum-like 
UV asymptotics~\cite{ope}, cf.\ \eqs\nr{rhn_asy}, \nr{dilepton_asy}, 
it is essential to choose parameters such that the masses appearing in 
the ``subleading'' terms in the Hamiltonian, 
  $ (m_{b}^2 - m_{a}^2 - \MM)^2/k^2 $ and 
  $ (m_{a}^2 - m_{b}^2 - \MM)^2/k^2 $,  
match those appearing in
the matrix elements squared, $\widetilde\Theta$.
Hence, in the case of $\widetilde \Theta^\mp_{ }$
of \eqs\nr{ThetaM_new} and \nr{ThetaP_new}, respectively, 
one should use $m_a^2=0$, $m_b^2=m_{\aS\T}^2$ 
in the Hamiltonian of \eq\nr{H_inter}. 
The dilepton Hamiltonian is instead unambiguous, 
given that $m_q^2-m_{\bar q}^2= m_{q \T}^2-m_{\bar q \T}^2=0$.

%%%%%%%%%%%%%%%%%%%%%%%%%%% SECTION %%%%%%%%%%%%%%%%%%%%%%%%%%%%%%%%%%%
%
\section{Matching of IR divergences}
\la{se:subtraction}

Apart from the 
$1+n \leftrightarrow 2+n $ scatterings that LPM resummation deals with, 
physics problems normally involve
$2\leftrightarrow 2$ and $1\leftrightarrow 3$ scatterings as well as
virtual corrections to $1\leftrightarrow 2$ scatterings that cancel
their IR divergences. However, the thermal part of the virtual 
corrections to $1\leftrightarrow 2$ scatterings also includes 
the HTL effects that lead to the thermal masses appearing 
in the LPM resummation. These processes must not be counted twice, 
and therefore a subtraction is needed. Specifically, the subtraction
should remove the most IR sensitive contributions, enhanced by
$\sim g^2T^2/M^2$ over the Born-level result when $M \to 0$, 
from the virtual corrections to $1\leftrightarrow 2$ scatterings, 
as these now appear as 
a part of LPM resummation,  schematically as 
$\sim \sum_{n=0}^{\infty}(g^2T^2/M^2)^n_{ }$. 

Depending on the formalism, 
the subtraction can be implemented either on the side 
of the virtual corrections to $1\leftrightarrow 2$ 
scatterings~\cite{interpolation}, or on the 
side of LPM resummation~\cite{dilepton}. 
The subtraction offers 
for a crosscheck of LPM resummation itself, 
verifying that 
IR sensitive effects that lead to a powerlike breakdown 
of the naive perturbative series are matched. 

To implement the subtraction on the side of LPM result, 
we should ``re-expand'' the latter to $\rmO(g^2T^2)$. 
Given that the width in 
\eqs\nr{H}, \nr{phi} is of $\rmO(g^4)$ if we formally expand 
in $\mE^2$, the only contribution at $\rmO(g^2T^2)$ comes from 
the thermal masses in \eq\nr{H_inter}. Therefore, 
we can now focus on \eq\nr{1to2_subtr0}, {\it viz.}\  
\be
 \Gamma^\rmi{inter}_{1\leftrightarrow 2}
 = 
% \!\! & = & \!\! 
 \int_{\epsilon_{a\T}^-}^{\epsilon_{a\T}^+} 
 \! {\rm d}\epsilon^{ }_a \, 
 \frac{ 
 \sign({\epsilon^{ }_a(\omega - \epsilon^{ }_a)})
 }{16\pi k}
 \,
     \bigl[1+ n^{ }_{\sigma_a}(\epsilon^{ }_a - \mu^{ }_a)
            + n^{ }_{\sigma_b}(\omega - \epsilon^{ }_a - \mu^{ }_b) \bigr]
 \Theta\bigl[\,\epsilon^{ }_a,p_{\perp\T}^2(\epsilon^{ }_a)\,\bigr] 
 \;, \la{1to2_subtr1}  
\ee
where 
$p_{\perp\T}^2(\epsilon_{a\T}^{\pm}) = 0$
at the integration boundaries.

If we now set $m^2_{a\T}\to m^2_a$, $m^2_{b\T}\to m_b^2$, we recover 
the Born result: 
\be
 \Gamma^\rmi{inter}_{1\leftrightarrow 2} |^{(g^0)}_{ }
 = 
 \Gamma^\rmi{Born}_{1\leftrightarrow 2}
 \;. \la{subtract_lo}
\ee
The first correction reads 
\be
 \Gamma^\rmi{inter}_{1\leftrightarrow 2} |^{(g^2)}_{ }
 = 
 \delta m_{a\T}^2 
 \frac{ \partial \Gamma^\rmi{inter}_{1\leftrightarrow 2} }
      { \partial m_{a\T}^2 }
 +  
 \delta m_{b\T}^2 
 \frac{ \partial \Gamma^\rmi{inter}_{1\leftrightarrow 2} }
      { \partial m_{b\T}^2 }
 \;.
\ee
% where
% $
%  \delta m_{a\T}^2 \equiv m_{a\T}^2 - m_a^2
% $.
Straightforward differentiation yields
\ba
 && \hspace*{-1.5cm} 
 \frac{ \partial \Gamma^\rmi{inter}_{1\leftrightarrow 2} }
      { \partial m_{a\T}^2 }
 \; = \;  
 \frac{ \sign({\epsilon^{+}_a(\omega - \epsilon^{+}_a)}) }{16\pi k}
 \,
     \bigl[1+ n^{ }_{\sigma_a}(\epsilon^{+}_a - \mu^{ }_a)
            + n^{ }_{\sigma_b}(\omega - \epsilon^{+}_a - \mu^{ }_b) \bigr]
 \Theta(\epsilon^{+}_a, 0 ) 
 \frac{\partial \epsilon_{a\T}^+}{\partial m_{a\T}^2 }
%%%%%% 
 \nn 
 & - &  
 \frac{ \sign({\epsilon^{-}_a(\omega - \epsilon^{-}_a)}) }{16\pi k}
 \,
     \bigl[1+ n^{ }_{\sigma_a}(\epsilon^{-}_a - \mu^{ }_a)
            + n^{ }_{\sigma_b}(\omega - \epsilon^{-}_a - \mu^{ }_b) \bigr]
 \Theta( \epsilon^{-}_a, 0 ) 
 \frac{\partial \epsilon_{a\T}^-}{\partial m_{a\T}^2 }
 \la{subtract}  
%%%%%%
 \\
 & + &
 \int_{\epsilon_a^-}^{\epsilon_a^+} 
 \! {\rm d}\epsilon^{ }_a \, 
 \frac{ 
 \sign({\epsilon^{ }_a(\omega - \epsilon^{ }_a)})
 }{16\pi k}
 \,
     \bigl[1+ n^{ }_{\sigma_a}(\epsilon^{ }_a - \mu^{ }_a)
            + n^{ }_{\sigma_b}(\omega - \epsilon^{ }_a - \mu^{ }_b) \bigr]
 \frac{\partial\Theta( \epsilon^{ }_a,p_\perp^2 ) } 
      {\partial p_\perp^2}
 \frac{\partial p_{\perp\T}^2}{\partial m_{a\T}^2 }
 \;, \nonumber 
\ea
and similarly for 
$
 { \partial \Gamma^\rmi{inter}_{1\leftrightarrow 2} } / 
      { \partial m_{b\T}^2 }
$.
The derivatives originate from the Dirac-$\delta$ in \eq\nr{1to2_subtr0}, 
\ba
 \frac{\partial \epsilon_{a\T}^{\pm}}{\partial m_{a\T}^2}
 & = &  
 \mp
   \frac{k^2 ( \omega - \epsilon^{\pm}_a )  }
     {\omega M^2 ( \epsilon_a^+ - \epsilon_a^- ) }
 \;, \quad
 \frac{\partial \epsilon_{a\T}^{\pm}}{\partial m_{b\T}^2}
 \; = \;
 \mp
    \frac{k^2 \epsilon_a^{\pm} }
   {\omega M^2 ( \epsilon_a^+ - \epsilon_a^- )}
 \;, \la{ir_boundaries} \\ 
%%%%
 \frac{\partial p_{\perp\T}^2}{\partial m_{a\T}^2}
 & = &  - \frac{ \omega - \epsilon^{ }_a }{\omega}
 \;, \quad 
 \frac{\partial p_{\perp\T}^2}{\partial m_{b\T}^2}
 \; = \; - \frac{\epsilon^{ }_a}{\omega}
 \;. \la{ir_integrand}
\ea
% where we have denoted 
% $\epsilon^{ }_b \equiv \omega - \epsilon^{ }_a$
% and
% $\epsilon^{\pm}_b \equiv \omega - \epsilon^{\pm}_a$.

Now, going to the UR limit, the integration boundaries 
from \eq\nr{epm} take the values 
$
 \epsilon_a^{\pm} 
 \stackrel{\rmiii{UR}}{\longrightarrow} 
 \{-\infty, 0,\omega,+\infty \}
$.
IR divergences could therefore originate from inverse powers of
$
  \epsilon_a^{\pm} 
$
or 
$
  \omega - \epsilon_a^{\pm} 
$
on the first two rows of \eq\nr{subtract}, or from inverse powers of 
$
 \epsilon_a^{ }
$ 
or 
$
 \omega - \epsilon_a^{ }
$
on the third row of \eq\nr{subtract}.
{}From \eq\nr{ir_boundaries}, 
we see that no inverse powers originate from
the boundary terms, however the division by $\MM$ shows that 
contributions 
$
 \rmO(\delta m_{a\T}^2/\MM)
$,
$
 \rmO(\delta m_{b\T}^2/\MM)
$
do appear. 
{}In \eq\nr{ir_integrand}, we see no inverse powers either. 
However, virtual corrections do contain inverse powers that
take the form of the last line of \eq\nr{subtract}. 
Therefore, it is essential to have the correct 
representation for $\Theta$, such that 
$
 \partial^{ }_{p_\perp^2}\Theta(\epsilon^{ }_a,p_\perp^2)
$
reproduces these IR divergences. 

Let us illustrate the matching of IR divergences for the 
case of right-handed neutrinos  first. 
A method to determine 
virtual corrections was worked out in ref.~\cite{phasespace}. The starting
point is to determine the real $1\rightarrow 3$ rate, 
given in its \eq(2.13). 
Identifying the poles and residues of the matrix element squared, the
virtual corrections can be identified, 
given in \eq(2.37) of ref.~\cite{phasespace}. 
The virtual corrections contain thermal
1-loop integrals of various types, of which only a subclass can 
lead to a HTL 
contribution, proportional to $T^2$~\cite{ht3}.
Among those in \eq(2.37) of ref.~\cite{phasespace}, 
the only one is the mixed 
fermion-boson loop, weighted by a loop momentum, denoted by 
$
             B(\P^{ }_{\bla} \,;\, \ala,\aQ)
            \, \E\cdot\P^{ }_{\ala}
$.
Specifically, carrying out the angular integral in \eq(2.29) of 
ref.~\cite{phasespace} and taking the UR limit, yields
($\tilde{\P} = (\tilde{\epsilon},\tilde{\vec{p}})$,
$\tilde{p} = |\tilde{\vec{p}}|$)
\ba
 && \hspace*{-1.0cm}
 B(\tilde\P;a,b) (\alpha \P^{ }_a + \beta \P^{ }_b)
%%%%%
 \nn[2mm] 
 &   \stackrel{\rmiii{UR}}{\longrightarrow}  & 
 \biggl( 0, \frac{\tilde{\vec{p}}}{{\tilde p}^2}\biggr)
 \frac{  \alpha - \beta }{8\pi^2}
 \int_0^\infty \! {\rm d}\epsilon \, \epsilon \, 
 \bigl[
    n^{ }_{\sigma_a}(\epsilon^{ } - \mu^{ }_a )  
  + n^{ }_{\sigma_a}(\epsilon^{ } + \mu^{ }_a )  
  - n^{ }_{\sigma_b}(\epsilon^{ } - \mu^{ }_b )  
  - n^{ }_{\sigma_b}(\epsilon^{ } + \mu^{ }_b )  
 \bigr] 
%%%%%
 \nn 
 & + & 
 \biggl( 1, \frac{ {\tilde{\vec{p}}} \, {\tilde{\epsilon}} } 
                 { {\tilde{p}}^2                         }\biggr) 
 \rmO\biggl( \frac{T^2,\mu_i^2}{\tilde{p}} \biggr)
 \;. \la{B_T}
\ea
If one of the particles is a fermion and the other is a boson, 
the thermal distributions add up 
($n^{ }_- = - \nF^{ }$ according to \eq\nr{n_sigma}), and we recover
a thermal fermion mass squared, notably
\be
 \frac{ 1 }{8\pi^2}
 \int_0^\infty \! {\rm d}\epsilon \, \epsilon \, 
 \bigl[
    \nF^{ }(\epsilon^{ } - \mu^{ }_a )  
  + \nF^{ }(\epsilon^{ } + \mu^{ }_a )  
  + 2 \nB^{ }(\epsilon^{ } )  
 \bigr] 
 = 
 \frac{1}{16}
 \biggl(
   T^2 + \frac{\mu_a^2}{\pi^2} 
 \biggr) 
 \;. \la{mass_T}
\ee

Applying \eqs\nr{B_T}, \nr{mass_T} to the appropriate term in \eq(2.37)
of ref.~\cite{phasespace}, we find
\ba
% && \hspace*{-1.5cm} 
 \Delta \Gamma^{\rmi{Born}(\tau)}_{1 \leftrightarrow 2}
 & \supset & 
 - 2 ( g_1^2 + 3 g_2^2 )\,
  \scat{1\leftrightarrow2}(\bla,\aS)
            \, B(\P^{ }_{\bla} \,;\, \ala,\aQ)
            \, \E^{\tau}_{ }\cdot\P^{ }_{\ala}
%%%%%
 \la{virtual_rhn_0} \\[2mm] 
 & \stackrel{\rmiii{UR}}{\approx} & 
  2 \,
 \scat{1\leftrightarrow2}(\bla,\aS) \, 
 \E^{\tau}_{ }\cdot 
 \biggl[ 
 \frac{ \delta m^2_{\ala\T} }{ \epsilon^{ }_{\bla} } \, 
 \biggl( 0, \frac{\vec{p}^{ }_{\bla}}{\epsilon^{ }_{\bla}} \biggr)
 + 
 \rmO\biggl( \frac{g^2T^2,g^2\mu_{\ala}^2}{\epsilon^{ }_{\bla}} \biggr) \, 
 \biggl( 1, \frac{\vec{p}^{ }_{\bla}}{\epsilon^{ }_{\bla}} \biggr)
 \biggr]
 \;. \hspace*{6mm} \la{virtual_rhn_1}
\ea
Here $\bla$ labels an on-shell lepton, distinguished
from the $\ala$ inside the loop $B$, whereas the vectors 
$
 \E^{\tau}_{ }
$
are from \eq\nr{E_T}.
We note that $\E^{-}_{ }$ is of $\rmO(M^2/\omega)$ in the UR limit, 
whereby $ \Delta \Gamma^{\rmii{Born$(-)$}}_\rmii{$1 \!\leftrightarrow\! 2$}$ 
is of $\rmO(g^4 T^2)$ and beyond our resolution. In contrast, 
$\E^{+}_{ } \stackrel{\rmiii{UR}}{\approx}(\omega,\vec{k})$ 
is of $\rmO(\omega)$. Recalling from \eq\nr{parallel} that 
$
 \vec{k}\cdot\vec{p}^{ }_{\bla} 
 \stackrel{\rmiii{UR}}{\approx}
 \omega \epsilon^{ }_{\bla}
$, 
the 2nd term in the square brackets in \eq\nr{virtual_rhn_1} 
drops out for $\tau = +$. 
In total, then, 
\be
 \Delta \Gamma^{\rmi{Born}(+)}_{1 \leftrightarrow 2}
 \stackrel{\rmiii{UR}}{\supset}
 -2 \, \delta m^2_{\ala\T} \,
 \scat{1\leftrightarrow2}(\bla,\aS)\,
 \frac{\omega}{\epsilon^{ }_{\bla}}
 \;. \la{virtual_rhn_2}
\ee
This shows an IR divergence 
(inverse power of $\epsilon^{ }_{\bla}$)
of the type that appears on the last
line of \eq\nr{subtract}. 
Recalling from \eq\nr{ThetaP_htl} that 
$
 \Theta^{+}_{ } 
 \stackrel{\rmiii{UR}}{\approx}
 2 \omega p_\perp^2 / \epsilon^{ }_\ala
$, 
\eq\nr{ir_integrand} 
indicates that 
\eq\nr{subtract} 
exactly matches 
\eq\nr{virtual_rhn_2} 
at $\epsilon^{ }_{\bla}\to 0$, 
if we have set $m_{a\T}^2 = \delta m_{\ala\T}^2$. 

The same exercise can be carried out for the dilepton case. 
In the normalization employed in \eq\nr{Gamma_munu}, 
the would-be $1\rightarrow 3$ contribution to the interaction rate reads
\ba
 && \hspace*{-1.0cm}
 \Gamma^{\rmi{Born}(\mu\nu)}_{1\to 3}
 \; = \; 
 \scat{1\to 3}(q,\ag,\bar{q}) \, 8 g_3^2 \CF \Nc^{ } \, \biggl\{
 \,  \eta^{\mu\nu}_{ }
%%%%
 \nn 
 & + & 
 \frac{
       \eta^{\mu\nu}_{ }(\s{q\bar{q}}^{ } + \MM)
       - 2 \bigl( \P^{\mu}_q \K^{\nu}_{ } + \P^{\nu}_q \K^{\mu}_{ }\bigr) 
       }
      {2(\s{q\ag}^{ } - m_q^2)}
 + 
 \frac{
       \eta^{\mu\nu}_{ }(\s{q\bar{q}}^{ } + \MM)
       - 2 \bigl( \P^{\mu}_{\bar{q}} \K^{\nu}_{ }
                + \P^{\nu}_{\bar{q}} \K^{\mu}_{ }\bigr) 
       }
      {2(\s{\bar{q}\ag}^{ } - m_q^2)}
%%%%
 \nn 
 & + & 
 \frac{ m_q^2 \bigl[ 
       \eta^{\mu\nu}_{ } \MM 
       - 2 \bigl( \P^{\mu}_{\bar{q}} \K^{\nu}_{ }
                + \P^{\nu}_{\bar{q}} \K^{\mu}_{ }\bigr) 
       + 4 \P^{\mu}_{\bar{q}} \P^{\nu}_{\bar{q}}
       \bigr] }
      {\bigl( \s{q\ag}^{ } - m_q^2 \bigr)^2 }
%%%%
 \nn 
 & + & 
 \frac{ m_q^2 \bigl[ 
       \eta^{\mu\nu}_{ } \MM 
       - 2 \bigl( \P^{\mu}_{{q}} \K^{\nu}_{ }
                + \P^{\nu}_{{q}} \K^{\mu}_{ }\bigr) 
       + 4 \P^{\mu}_{{q}} \P^{\nu}_{{q}}
       \bigr] }
      {\bigl( \s{\bar{q}\ag}^{ } - m_q^2 \bigr)^2 }
 \la{scat1to3_dilepton}
%%%%
 \\ 
 & + & 
 \frac{ (2 m_q^2 - \MM)  \bigl[ 
       \eta^{\mu\nu}_{ } \MM 
       -   \bigl( \P^{\mu}_{\ag} \K^{\nu}_{ }
                + \P^{\nu}_{\ag} \K^{\mu}_{ }
           \bigr) 
       - 2 \bigl( 
                  \P^{\mu}_{{q}} \P^{\nu}_{\bar{q}}
                + \P^{\nu}_{{q}} \P^{\mu}_{\bar{q}}
           \bigr)
          \bigr] 
        - 2 \MM \P^{\mu}_{\ag} \P^{\nu}_{\ag}
        }
      {
           \bigl( \s{{q}\ag}^{ } - m_q^2 \bigr)
           \bigl( \s{\bar{q}\ag}^{ } - m_q^2 \bigr) }
%%%%
 \,\biggr\} 
 \;. 
 \nonumber 
\ea
Here $g$ denotes a gluon, 
$\s{ab}^{ } \equiv (\P^{ }_a + \P^{ }_b)^2$, 
and $\scat{1\to 3}$ is defined in \eqs(2.2)--(2.4) of ref.~\cite{phasespace}.

Like above, virtual corrections can be deduced from the poles
of \eq\nr{scat1to3_dilepton}, 
and only first-order poles lead to terms involving 
HTLs in the UR limit. It is helpful to 
consider the transverse and vector projections, from 
\eqs\nr{GammaT} and \nr{GammaV}. Both yield the same HTLs, 
which after writing $\s{q\bar{q}}^{ } = (\K - \P^{ }_{\ag})^2$  
gives an expression analogous to \eq\nr{virtual_rhn_0}, {\it viz.}\ 
\be
 \Delta \Gamma^{\rmi{Born}(\rmii{T},\rmii{V})}_{1 \leftrightarrow 2}
 \; \supset \; 
  16 g_3^2 \CF \Nc^{ }\, 
  \bigl[ 
              \scat{1\leftrightarrow2}(\tilde{q},\bar{q})
            \, B(\P^{ }_{\tilde{q}} \,;\, {q},\ag) 
        + 
              \scat{1\leftrightarrow2}({q},\tilde{\bar{q}})
            \, B(\P^{ }_{\tilde{\bar{q}}} \,;\, \bar{q},\ag) 
  \bigr] 
            \, \K^{ }_{ }\cdot\P^{ }_{\ag}
 \;. \la{virtual_dilepton_1}
\ee
Given that T and V yield the same IR divergence, 
there is none in the L channel. 

In the UR limit, \eq\nr{virtual_dilepton_1} can be approximated like in
\eq\nr{virtual_rhn_1}. Again the 2nd term in the square brackets
drops out, when we make use of 
$
 \vec{k}\cdot\vec{p}^{ }_{\tilde{q}} 
 \stackrel{\rmiii{UR}}{\approx} 
 \omega \epsilon^{ }_{\tilde{q}}
$
and 
$
 \vec{k}\cdot\vec{p}^{ }_{\tilde{\bar{q}}} 
 \stackrel{\rmiii{UR}}{\approx} 
 \omega \epsilon^{ }_{\tilde{\bar{q}}}
$.
This leads to the analogue of \eq\nr{virtual_rhn_2}, 
{\it viz.}\  
\be
 \Delta \Gamma^{\rmi{Born}(\rmii{T},\rmii{V})}_{1 \leftrightarrow 2}
 \stackrel{\rmiii{UR}}{\supset}
 -4\Nc^{ } \, m_{\infty}^2 \, \biggl[ 
 \scat{1\leftrightarrow2}(\tilde{q},\bar{q})\,
 \frac{\omega}{\epsilon^{ }_{\tilde{q}}}
 + 
 \scat{1\leftrightarrow2}({q},\tilde{\bar{q}})\,
 \frac{\omega}{\epsilon^{ }_{\tilde{\bar{q}}}}
 \biggr]
 \;. \la{virtual_dilepton_2}
\ee

Let us compare \eq\nr{virtual_dilepton_2} with \eq\nr{subtract}. 
{}From \eq\nr{ThetaT} and from the projection of \eq\nr{Gamma_munu}
according to \eq\nr{GammaV}, we have 
\ba
 \Gamma^{\rmi{Born}(\rmii{T})}_{1\leftrightarrow 2} 
 & = &   
 \scat{1\leftrightarrow 2}(q,\bar{q}) 
 \, 
  4\Nc^{ }
 \bigl(
   - 2 p_\perp^2 + M^2
 \bigr)
 \;, \\ 
%%%%%
 \Gamma^{\rmi{Born}(\rmii{V})}_{1\leftrightarrow 2} 
 & = &   
 \scat{1\leftrightarrow 2}(q,\bar{q}) 
 \, 
  4\Nc^{ }
 \bigl(
  \, 2 m_q^2 + M^2
 \bigr)
 \;.
\ea
As discussed below \eq\nr{ir_integrand}, $p_\perp^2$ as such does not
lead to an IR divergence. But if we eliminate $M^2$ in favour of $p_\perp^2$
by making use of \eq\nr{1to2_step2}, 
as has been done in order to arrive at \eq\nr{ThetaT_new}, 
then the corresponding contributions 
depend on the thermal mass correction as 
\be
 \frac{ \partial M^2 }{ \partial m_\infty^2  } 
 \; \equiv \;  
 \frac{\partial M^2}{\partial p_\perp^2 } \,
 \biggl(
   \frac{\partial p_{\perp\T}^2 }{\partial m_{q\T}^2} 
  + 
   \frac{\partial p_{\perp\T}^2 }{\partial m_{\bar{q}\T}^2} 
 \biggr) 
 \; \stackrel{\rmiii{UR}}{\approx} \; 
 \frac{\omega^2}{\epsilon^{ }_{q}\epsilon^{ }_{\bar{q}} }
 \biggl(
  - \frac{\epsilon^{ }_{\bar{q}}}{\omega}
  - \frac{\epsilon^{ }_{q}}{\omega}
 \biggr)
 \; = \;
 - 
 \biggl( 
   \frac{\omega}{\epsilon^{ }_{q}} 
 + 
   \frac{\omega}{\epsilon^{ }_{\bar{q}}}
 \biggr)
 \;. \hspace*{5mm}
\ee
Thereby \eq\nr{virtual_dilepton_2} is indeed reproduced. 
Incidentally, this is the same divergence that renders the 
strict NLO expression for the 
vector spectral function logarithmically
divergent and discontinuous across the light cone~\cite{harvey}. 

To summarize this section, we have verified that after a proper 
choice of variables in the matrix elements squared, 
the LPM resummed results
match the IR divergences that appear in 
virtual corrections to the Born result. 
If an LPM resummed result is combined with a computation
which already includes virtual corrections to $1\leftrightarrow 2$
scatterings, \eqs\nr{subtract_lo}--\nr{subtract} should be subtracted, 
in order to avoid double counting. 

%%%%%%%%%%%%%%%%%%%%%%%%%%% SECTION %%%%%%%%%%%%%%%%%%%%%%%%%%%%%%%%%%%
%
\section{Numerical evaluation}
\la{se:numerics}

%%%%%%%%%%%%%%%%%%%%%%%%%%%%%%%%% FIGURE %%%%%%%%%%%%%%%%%%%%%%%%%%%%%%%%%
\begin{figure}[t]

\hspace*{-0.1cm}
\centerline{%
 \epsfysize=7.5cm\epsfbox{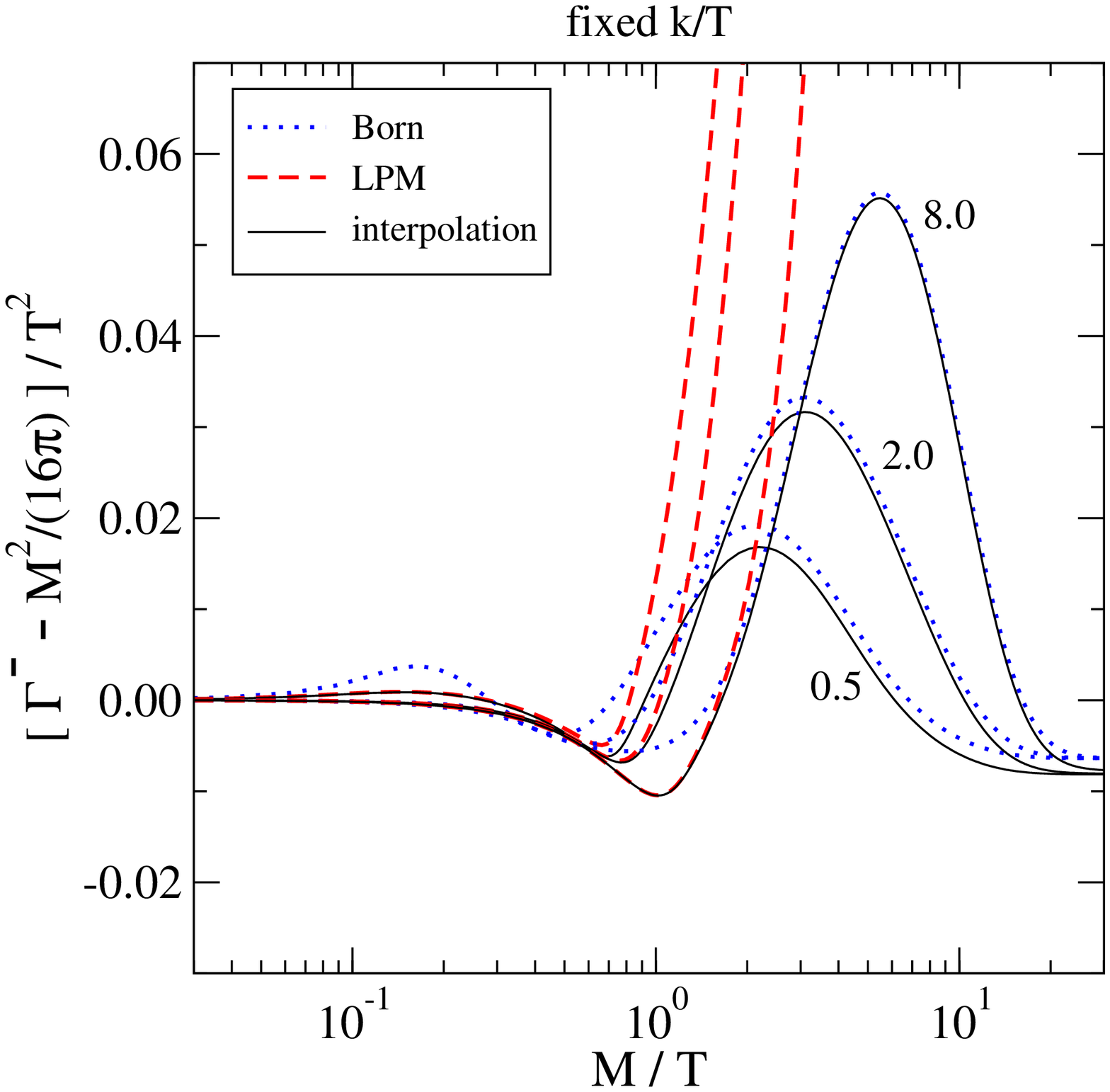}%
 \hspace{0.5cm}%
 \epsfysize=7.5cm\epsfbox{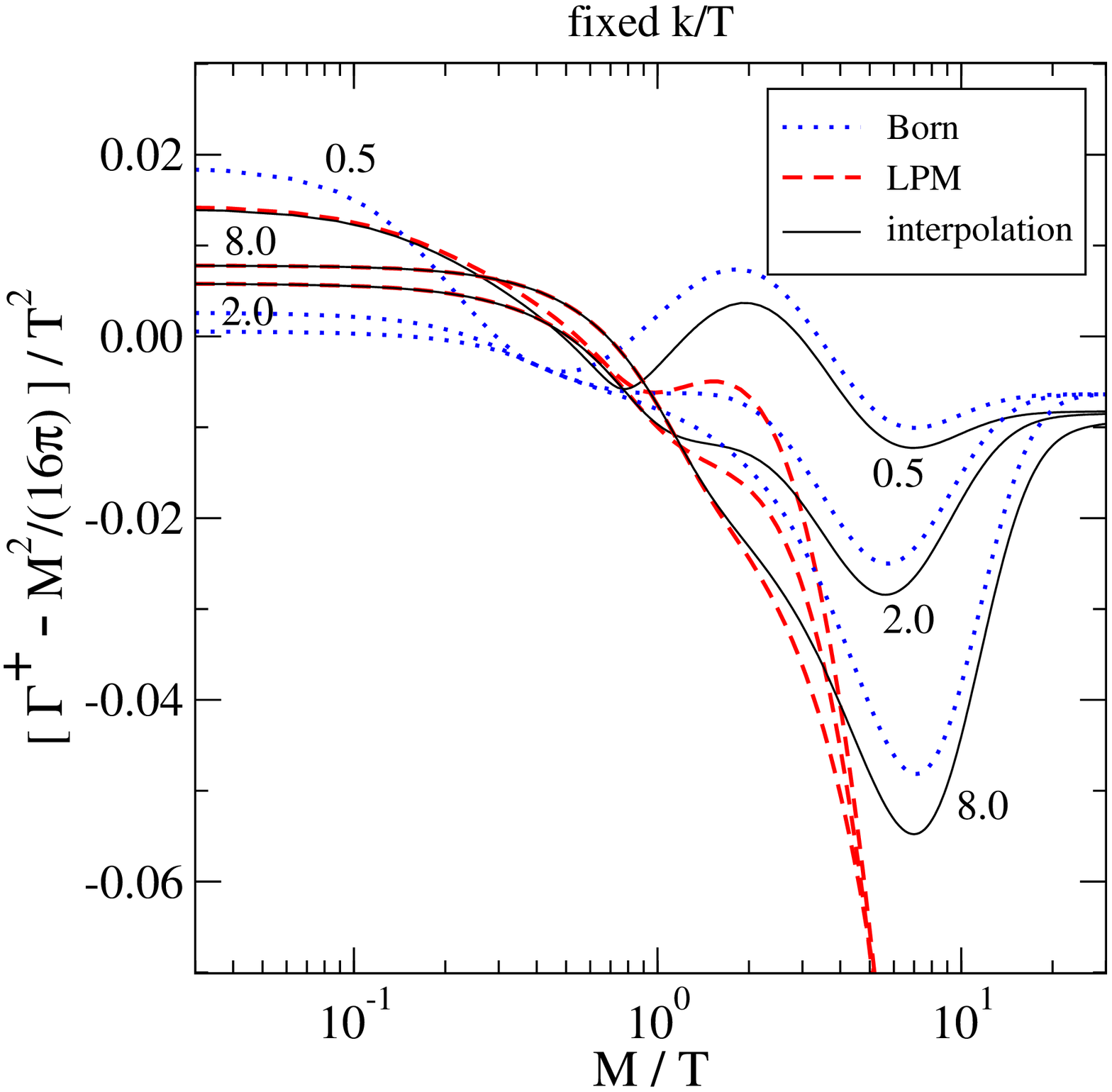}%
}

\caption[a]{\small
 Results for the right-handed neutrino interaction rates 
 in the $\tau = -$ (left)
 and $\tau = +$ (right) channels, after the subtraction of the 
 leading-order vacuum contributions at $M \gg m^{ }_{\aS\T}$,
 cf.\ \eq\nr{rhn_asy}. 
 The value of $k/T$ has been indicated next to the curves. 
}

\la{fig:RHN}
\end{figure}
%%%%%%%%%%%%%%%%%%%%%%%%%%%%%%%%%%%%%%%%%%%%%%%%%%%%%%%%%%%%%%%%%%%%%%%%%%%

%%%%%%%%%%%%%%%%%%%%%%%%%%%%%%%%% FIGURE %%%%%%%%%%%%%%%%%%%%%%%%%%%%%%%%%
\begin{figure}[t]

\hspace*{-0.1cm}
\centerline{%
 \epsfysize=7.5cm\epsfbox{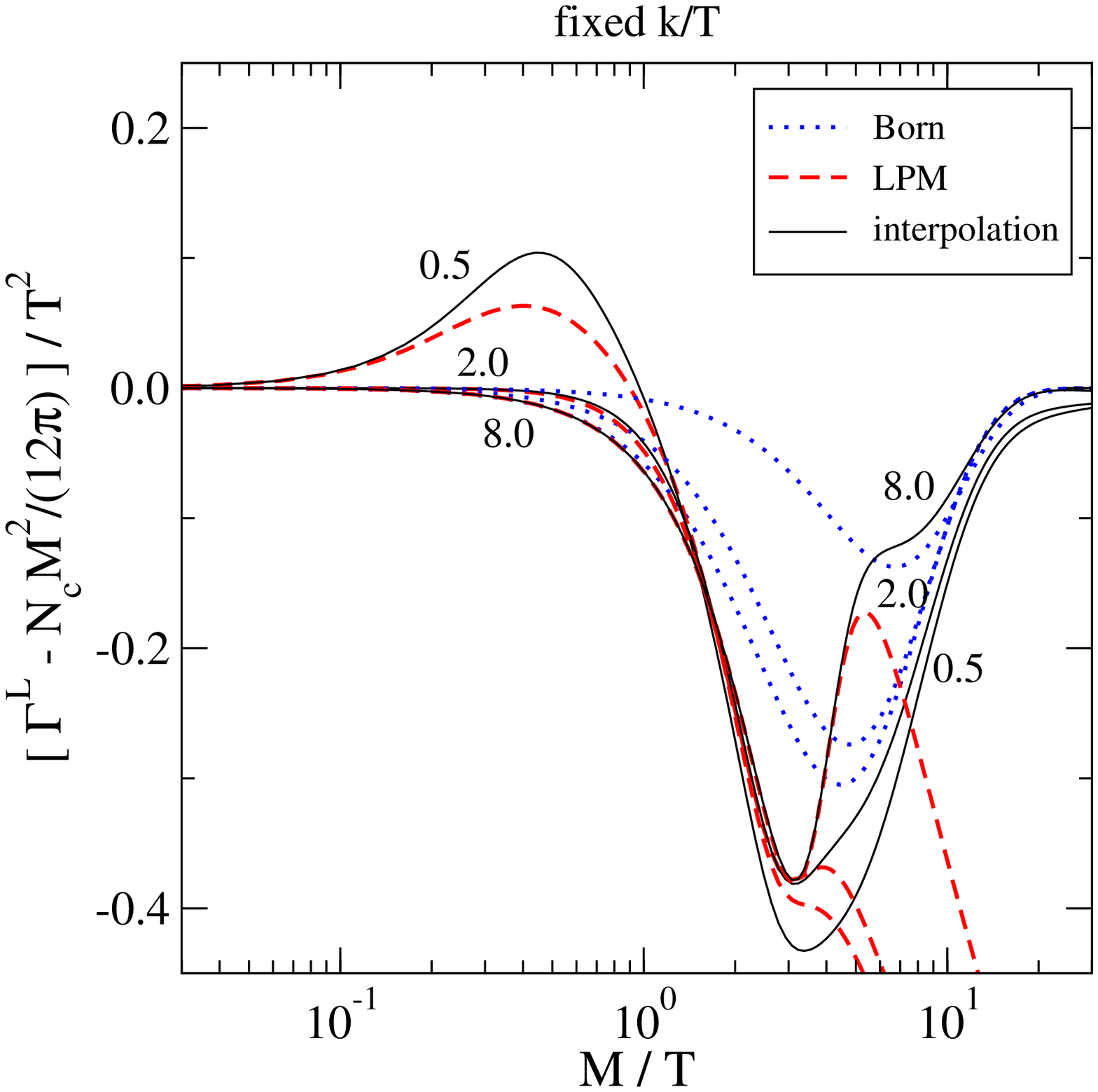}%
 \hspace{0.5cm}%
 \epsfysize=7.5cm\epsfbox{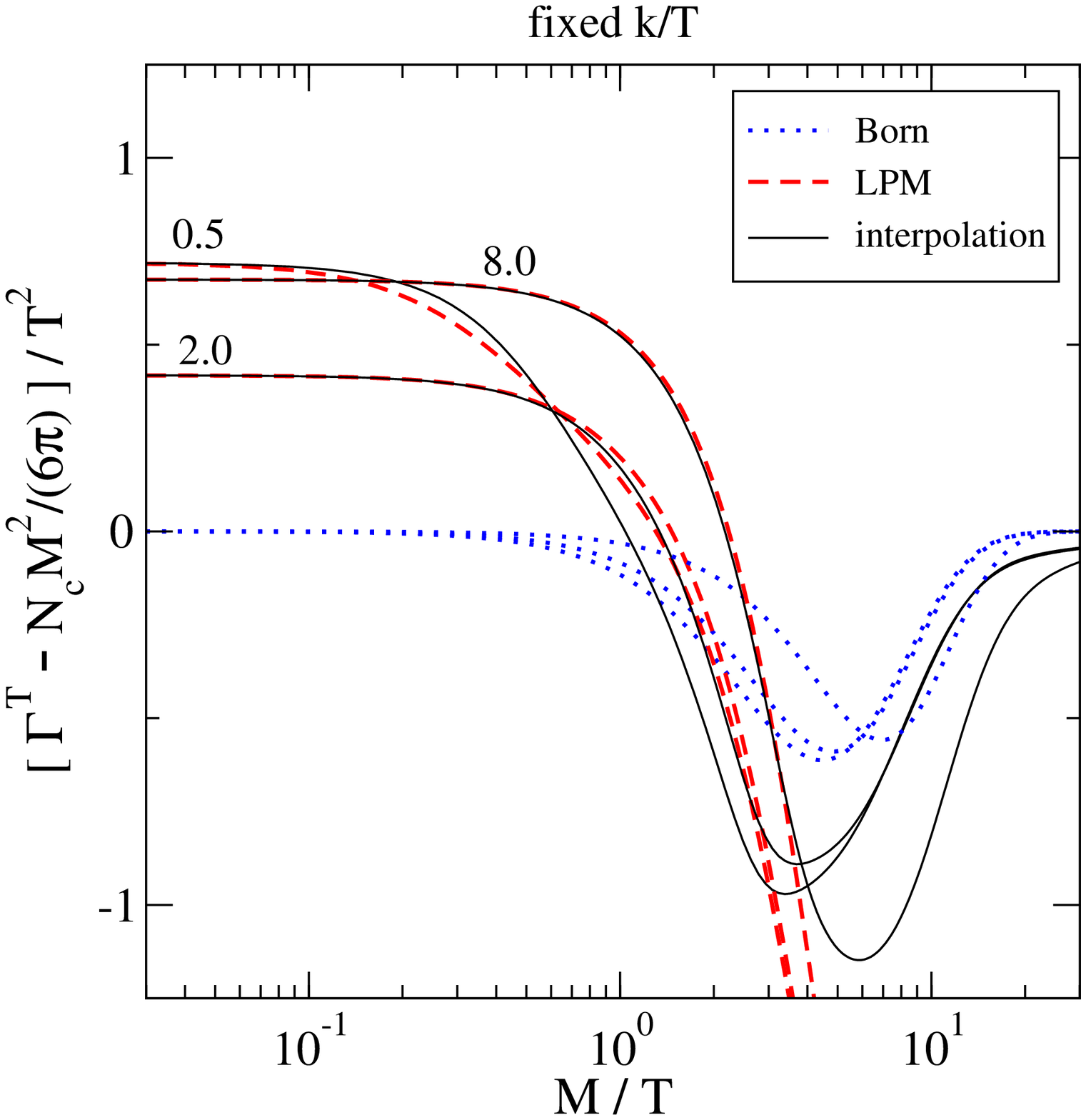}%
}

\caption[a]{\small
 Results for the dilepton interaction rates in the L (left)
 and T (right) channels, after the subtraction of the 
 leading-order vacuum contributions, 
 cf.\ \eq\nr{dilepton_asy}.
 The value of $k/T$ has been indicated next to the curves. 
}

\la{fig:dilepton}
\end{figure}
%%%%%%%%%%%%%%%%%%%%%%%%%%%%%%%%%%%%%%%%%%%%%%%%%%%%%%%%%%%%%%%%%%%%%%%%%%%

In this section, we illustrate the recipe of 
\se\ref{se:interpolation} numerically. 
In \fig\ref{fig:RHN}, results based on
$
 \Gamma^\rmi{Born}_{1\leftrightarrow 2}
$, 
$
 \Gamma^\rmii{LPM}_{1+n\leftrightarrow 2+n} 
$, 
$
 \Gamma^\rmi{inter}_{1+n\leftrightarrow 2+n}
$,
are shown for right-handed neutrinos of either helicity, 
with matrix elements squared taken from \eqs\nr{ThetaM_new} and
\nr{ThetaP_new}. 
As we are in the symmetric phase, 
the lepton vacuum mass has been omitted and the scalar vacuum mass
replaced by its thermal counterpart
($m^{ }_{\ala} = 0$, $m^{ }_{\aS}\to m^{ }_{\aS\T}$); 
the thermal masses have been set to 
$\delta m^{ }_{\ala\T} = 0.3T$, $m^{ }_{\aS\T} = 0.4T$, 
$m^{ }_{\rmii{E1}} = 0.5T$, $m^{ }_{\rmii{E2}} = 0.9T$; 
and the couplings to $g^{ }_1 = 1/3$, $g^{ }_2 = 2/3$.
A few separate values of $k/T$ have been chosen, 
with the results plotted as a function of $M/T$.  
The same exercise is repeated for 
both channels of the vector correlator
in \fig\ref{fig:dilepton},  
with matrix elements squared from \eqs\nr{ThetaL_new} and \nr{ThetaT_new},  
setting quark vacuum masses to zero ($m^{ }_q = 0$),
thermal masses to 
$m^{ }_\infty = 1.0T $, 
$m^{ }_{\rmii{E3}} = 2.2T$,  
and the gauge coupling to $\alphas^{ } = 0.3$.
For a better plotting resolution, 
leading-order vacuum contributions that would otherwise dominate
at large $M/T$ (cf.\ \se\ref{se:OPE}), have been subtracted in both systems. 

The conclusion from \figs\ref{fig:RHN} and \ref{fig:dilepton}
is that the interpolations indeed agree with LPM results for 
$M \ll \pi T$, and with the {\em shape} of the  
Born results for $M \gg \pi T$, 
however in the latter regime 
we observe constant offsets, cf.\ \se\ref{se:OPE}.
It is appropriate to stress 
that the offsets are only visible because we have 
subtracted the dominant terms, and that the offsets originate from 
fermionic thermal mass corrections which are absent 
in the Born results, 
and could be subtracted through a matching computation, 
as described in \se\ref{se:subtraction}. 
 
%%%%%%%%%%%%%%%%%%%%%%%%%%% SECTION %%%%%%%%%%%%%%%%%%%%%%%%%%%%%%%%%%%%%%

\section{Conclusions and outlook}
\la{se:concl}

At leading non-trivial order in the weak-coupling expansion, 
general thermal interaction rates consist 
of two classes of contributions.
On one hand, there are 
${2}\leftrightarrow{2}$ and the crossed
${1}\leftrightarrow{3}$ processes as well as 
virtual corrections to ${1}\leftrightarrow{2}$ processes
whose role is to cancel IR divergences. 
These have recently been analysed 
in some generality in ref.~\cite{phasespace}. 
On the other hand, there are 
real ${1+n}\leftrightarrow{2+n}$ processes, 
with $n\ge 0$, that need often to be summed to all orders in $n$, 
through LPM resummation~\cite{gelis3,amy1,agmz,bb1}. 
The latter set constitutes the focus of the present paper. 
Even though the two classes do not completely decouple from 
each other, in the sense that specific thermal corrections 
appearing in the virtual corrections to ${1}\leftrightarrow{2}$ 
processes also play a role in LPM resummation, whereby 
a subtraction procedure is needed for evading double counting 
(cf.\ \se\ref{se:subtraction}), for practical purposes 
the kinematics and the theoretical tools playing a role
in the two sets are very different. Therefore the methods
of ref.~\cite{phasespace} are of little help here, and 
a dedicated analysis has become necessary. 

Concretely, 
we have generalized the light-cone 
Hamiltonian that plays a role in LPM resummation, beyond the 
ultrarelativistic regime for which it was originally derived. 
Our key result is \eq\nr{H_inter}, which sets the Hamiltonian
in a form that respects the correct kinematics of the Born 
limit as well (cf.\ \eq\nr{1to2_step2}). 
 
Apart from the Hamiltonian, 
matrix elements squared, or splitting functions, play a role
in the formalism. Even though they can in principle 
be derived at Born level, 
the Born results can be put in various forms, 
by making use of energy-momentum conservation. Yet only 
specific representations reproduce the correct IR divergences
when the Hamiltonian is modified by thermal corrections. 
We have shown two separate ways to verify that
the variables chosen in \eqs\nr{ThetaM_new}--\nr{ThetaT_new}
are consistent from this point of view, 
going either through an explicit HTL computation 
(cf.\ \se\ref{ss:general}),
or through the inspection of the IR divergences that appear
in NLO virtual corrections to ${1}\leftrightarrow{2}$ 
processes (cf.\ \se\ref{se:subtraction}).

Another consistency check on the interpolation is obtained
by inspecting the UV domain, in which the mass of the probe
particle is made large. We have verified that our formalism
produces a qualitatively correct behaviour in this limit
(cf.\ \se\ref{se:OPE}).
For this, appearances of thermal masses need to be chosen consistently
in \eq\nr{H_inter} and in \eqs\nr{ThetaM_new}--\nr{ThetaT_new}.

Despite being able to cover a broad kinematic
regime with our interpolation, 
we stress that there is a particular corner, namely that
where momenta and masses are all small compared with the temperature, 
in which neither Born nor LPM results are reliable
(cf.\ \fig\ref{fig:overview}). Consequently our interpolation does 
not work in the vicinity of this domain. 
It would be interesting to remedy this shortcoming, 
however this represents a demanding task, 
due to the complicated resummations required. 

The computations that we have carried out apply formally 
at leading order in the weak-coupling expansion, i.e.\ 
$\rmO(g^2T^2)$ in the ultrarelativistic regime. A relevant question
is to what extent known or unknown NLO corrections, suppressed only
by $\rmO(g)$~\cite{sch}, 
could be incorporated in our setup. This would be straightforward 
for the objects appearing in \eq\nr{H_inter}, 
notably asymptotic masses~\cite{sch2,mass2} or the 
thermal width~\cite{width2,width3,width4,width5}.
Furthermore, in the dilepton case, 
an analytic continuation of our framework 
permits for the determination of certain
correlation lenghts, which may be compared with lattice
simulations, to test empirically the  
influence of NLO corrections~\cite{screening}. That said, 
there are also NLO effects whose inclusion requires 
a more significant effort, as they involve regions
where thermal fermions 
(active leptons or quarks in our examples) become soft. 
To treat these regions correctly, 
full-fledged HTL computations are required, 
generalizing on the ultrarelativistic approximation 
discussed in \se\ref{ss:general}. In addition, 
subtractions beyond those discussed in \se\ref{se:subtraction}
are needed, in order to avoid double counting
the soft-fermion region at first order
in the width~\cite{nlo_photon}.

%%%%%%%%%%%%%%%%%%%%%%%%%%% SECTION %%%%%%%%%%%%%%%%%%%%%%%%%%%%%%%%%%
%
\section*{Acknowledgements}

J.G.\ acknowledges support by the R\'egion Pays de la Loire under a PULSAR
grant. 
M.L.\ was supported by the Swiss National Science Foundation
(SNF), through grant 200020B-188712.

%%%%%%%%%%%%%%%%%%%%%%% APPENDIX %%%%%%%%%%%%%%%%%%%%%%%%%%%%%%%%%%%
%
\appendix
\renewcommand{\thesection}{\Alph{section}} % {Appendix~\Alph{section}}
\renewcommand{\thesubsection}{\Alph{section}.\arabic{subsection}}
\renewcommand{\theequation}{\Alph{section}.\arabic{equation}}
%
%%%%%%%%%%%%%%%%%%%%%%%%%%%%%%%%%%%%%%%%%%%%%%%%%%%%%%%%%%%%%%%%%%%%%%%%%%%

%%%%%%%%%%%%%%%%%%%%%%%% BIBLIO %%%%%%%%%%%%%%%%%%%%%%%%%%%%%%%%%%%%%%%%%%
%

\small{
 
}


\begin{thebibliography}{999}

\bibitem{gelis3}
  P.~Aurenche, F.~Gelis and H.~Zaraket,
  {\it Landau-Pomeranchuk-Migdal effect in thermal field theory,}
  Phys.\ Rev.\ D {62} (2000) 096012
  [hep-ph/0003326].
  %%CITATION = doi:10.1103/PhysRevD.62.096012;%%

\bibitem{amy1}
  P.B.~Arnold, G.D.~Moore and L.G.~Yaffe,
  {\it Photon emission from ultrarelativistic plasmas,}
  JHEP {11} (2001) 057
  [hep-ph/0109064].
  %%CITATION = doi:10.1088/1126-6708/2001/11/057;%%

\bibitem{agmz}
  P.~Aurenche, F.~Gelis, G.D.~Moore and H.~Zaraket,
  {\it Landau-Pomeranchuk-Migdal resummation for dilepton production,}
  JHEP {12} (2002) 006
  [hep-ph/0211036].
  %%CITATION = doi:10.1088/1126-6708/2002/12/006;%%

\bibitem{bb1}
  A.~Anisimov, D.~Besak and D.~B\"odeker,
  {\it Thermal production of relativistic Majorana neutrinos: 
  Strong enhancement by multiple soft scattering,}
  JCAP {03} (2011) 042
  [1012.3784].
  %%CITATION = doi:10.1088/1475-7516/2011/03/042;%%

\bibitem{app1}
  B.~Garbrecht, F.~Glowna and P.~Schwaller,
  {\it Scattering rates for leptogenesis:
  Damping of lepton flavour coherence and production of singlet neutrinos,}
  Nucl.\ Phys.\ B {877} (2013) 1
  [1303.5498].
  %%CITATION = doi:10.1016/j.nuclphysb.2013.08.020;%%

\bibitem{interpolation}
  I.~Ghisoiu and M.~Laine,
  {\it Right-handed neutrino production rate at $T > 160$ GeV,}
  JCAP {12} (2014) 032
  [1411.1765].
  %%CITATION = doi:10.1088/1475-7516/2014/12/032;%%

\bibitem{broken}
  J.~Ghiglieri and M.~Laine,
  {\it Neutrino dynamics below the electroweak crossover,}
  JCAP {07} (2016) 015
  [1605.07720].
  %%CITATION = doi:10.1088/1475-7516/2016/07/015;%%

\bibitem{app2}
  P.~Hern\'andez, M.~Kekic, J.~L\'opez-Pav\'on, J.~Racker and J.~Salvado,
  {\it Testable baryogenesis in seesaw models,}
  JHEP {08} (2016) 157
  [1606.06719].
  %%CITATION = doi:10.1007/JHEP08(2016)157;%%

\bibitem{degenerate}
  J.~Ghiglieri and M.~Laine,
  {\it Precision study of GeV-scale resonant leptogenesis,}
  JHEP {02} (2019) 014
  [1811.01971].
  %%CITATION = doi:10.1007/JHEP02(2019)014;%%

\bibitem{app3}
  D.~B\"odeker and D.~Schr\"oder,
  {\it Equilibration of right-handed electrons,}
  JCAP {05} (2019) 010
  [1902.07220].
  %%CITATION = doi:10.1088/1475-7516/2019/05/010;%%

\bibitem{app4}
  S.~Biondini and J.~Ghiglieri,
  {\it Freeze-in produced dark matter in the ultra-relativistic regime,}
  JCAP {03} (2021) 075
  [2012.09083].
  %%CITATION = doi:10.1088/1475-7516/2021/03/075;%%

\bibitem{app5}
  J.~Klari\'c, M.~Shaposhnikov and I.~Timiryasov,
  {\it Reconciling resonant leptogenesis and
  baryogenesis via neutrino oscillations,}
  Phys.\ Rev.\ D {104} (2021) 055010
  [2103.16545].
  %%CITATION = doi:10.1103/PhysRevD.104.055010;%%

\bibitem{app6}
  M.~Drees and B.~Najjari,
  {\it Energy Spectrum of Thermalizing High Energy Decay Products
  in the Early Universe,}
  2105.01935.
  %%CITATION = ;%%

\bibitem{nlo_dilepton}
  J.~Ghiglieri and G.D.~Moore,
  {\it Low mass thermal dilepton production at NLO
  in a weakly coupled quark-gluon plasma,}
  JHEP {12} (2014) 029
  [1410.4203].
  %%CITATION = doi:10.1007/JHEP12(2014)029;%%

\bibitem{harvey}
  G.~Jackson and M.~Laine,
  {\it Testing thermal photon and dilepton rates,}
  JHEP {11} (2019) 144
  [1910.09567].
  %%CITATION = doi:10.1007/JHEP11(2019)144;%%

\bibitem{cptheory}
  J.~Ghiglieri and M.~Laine,
  {\it GeV-scale hot sterile neutrino oscillations:
  a derivation of evolution equations,}
  JHEP {05} (2017) 132
  [1703.06087].
  %%CITATION = doi:10.1007/JHEP05(2017)132;%%

\bibitem{weldon} 
  H.A.~Weldon,
  {\it Effective fermion masses of order $gT$ 
  in high-temperature gauge theories
  with exact chiral invariance,}
  Phys.\ Rev.\ D {26} (1982) 2789.
  %%CITATION = doi:10.1103/PhysRevD.26.2789;%%

\bibitem{meg}
  M.E.~Carrington,
  {\it Effective potential at finite temperature in the Standard Model,}
  Phys.\ Rev.\ D {45} (1992) 2933.
  %%CITATION = doi:10.1103/PhysRevD.45.2933;%%

\bibitem{sum1}
  P.~Aurenche, F.~Gelis and H.~Zaraket,
  {\it A simple sum rule for the thermal gluon spectral 
  function and applications,}
  JHEP {05} (2002) 043
  [hep-ph/0204146].
  %%CITATION = doi:10.1088/1126-6708/2002/05/043;%%

\bibitem{sch}
  S.~Caron-Huot,
  {\it $O(g)$ plasma effects in jet quenching,}
  Phys.\ Rev.\ D {79} (2009) 065039
  [0811.1603].
  %%CITATION = doi:10.1103/PhysRevD.79.065039;%%

\bibitem{klimov}
  V.V.~Klimov,
  {\it Collective Excitations in a Hot Quark Gluon Plasma,}
  Sov.\ Phys.\ JETP {55} (1982) 199
  [Zh.\ Eksp.\ Teor.\ Fiz.\  {82} (1982) 336].
  %%CITATION = NONE;%% 

\bibitem{htl5}
  J.~Frenkel and J.C.~Taylor,
  {\it Hard thermal QCD, forward scattering and effective actions,}
  Nucl.\ Phys.\ B {374} (1992) 156.
  %%CITATION = doi:10.1016/0550-3213(92)90480-Y;%%
 
\bibitem{htl6}
  E.~Braaten and R.D.~Pisarski,
  {\it Simple effective Lagrangian for hard thermal loops,}
  Phys.\ Rev.\ D {45} (1992) 1827. 
  %%CITATION = doi:10.1103/PhysRevD.45.R1827;%%

\bibitem{agz}
  P.~Aurenche, F.~Gelis and H.~Zaraket,
  {\it Enhanced thermal production of hard dileptons by $3 \to 2$ processes,}
  JHEP {07} (2002) 063
  [hep-ph/0204145].
  %%CITATION = doi:10.1088/1126-6708/2002/07/063;%%

\bibitem{nlo_photon}
  J.~Ghiglieri, J.~Hong, A.~Kurkela, E.~Lu, G.D.~Moore and D.~Teaney,
  {\it Next-to-leading order thermal photon production
  in a weakly coupled quark-gluon plasma,}
  JHEP {05} (2013) 010
  [1302.5970].
  %%CITATION = doi:10.1007/JHEP05(2013)010;%%

\bibitem{ope}
  S.~Caron-Huot,
  {\it Asymptotics of thermal spectral functions,}
  Phys.\ Rev.\  D {79} (2009) 125009
  [0903.3958].
  %%CITATION = doi:10.1103/PhysRevD.79.125009;%%

\bibitem{salvio} 
  A.~Salvio, P.~Lodone and A.~Strumia,
  {\it Towards leptogenesis at NLO: 
  the right-handed neutrino interaction rate,}
  JHEP {08} (2011) 116
  [1106.2814].
  %%CITATION = doi:10.1007/JHEP08(2011)116;%%

\bibitem{nonrel}
  M.~Laine and Y.~Schr\"oder,
  {\it  Thermal right-handed neutrino production rate 
  in the non-relativistic regime,} 
  JHEP {02} (2012) 068
  [1112.1205].
  %%CITATION = doi:10.1007/JHEP02(2012)068;%%

\bibitem{biondini}
  S.~Biondini, N.~Brambilla, M.A.~Escobedo and A.~Vairo,
  {\it An effective field theory for non-relativistic Majorana neutrinos,}
  JHEP {12} (2013) 028
  [1307.7680].
  %%CITATION = doi:10.1007/JHEP12(2013)028;%%

\bibitem{dilepton}
  I.~Ghisoiu and M.~Laine,
  {\it Interpolation of hard and soft dilepton rates,}
  JHEP {10} (2014) 083
  [1407.7955].
  %%CITATION = doi:10.1007/JHEP10(2014)083;%%

\bibitem{phasespace}
  G.~Jackson and M.~Laine,
  {\it Efficient numerical integration of thermal interaction rates,}
  JHEP {09} (2021) 125
  [2107.07132].
  %%CITATION = doi:10.1007/JHEP09(2021)125;%%

\bibitem{ht3}
  E.~Braaten and R.D.~Pisarski,
  {\it Soft amplitudes in hot gauge theories: A general analysis,}
  Nucl.\ Phys.\ B {337} (1990) 569.
  %%CITATION = doi:10.1016/0550-3213(90)90508-B;%%
  
\bibitem{sch2}
  S.~Caron-Huot,
  {\it On supersymmetry at finite temperature,}
  Phys.\ Rev.\ D {79} (2009) 125002
  [0808.0155].
  %%CITATION = doi:10.1103/PhysRevD.79.125002;%%

\bibitem{mass2}
  G.D.~Moore and N.~Schlusser,
  {\it The nonperturbative contribution to asymptotic masses,}
  Phys.\ Rev.\ D {102} (2020) 094512
  [2009.06614].
  %%CITATION = doi:10.1103/PhysRevD.102.094512;%%

\bibitem{width2}
  M.~Panero, K.~Rummukainen and A.~Sch\"afer,
  {\it Lattice Study of the Jet Quenching Parameter,}
  Phys.\ Rev.\ Lett.\  {112} (2014)  162001
  [1307.5850].
  %%CITATION = doi:10.1103/PhysRevLett.112.162001;%%

\bibitem{width3}
  J.~Ghiglieri and H.~Kim,
  {\it Transverse momentum broadening and collinear radiation
  at NLO in the $\mathcal{N}=4$ SYM plasma,}
  JHEP {12} (2018) 049
  [1809.01349].
  %%CITATION = doi:10.1007/JHEP12(2018)049;%%

\bibitem{width4}
  G.~Jackson and M.~Laine,
  {\it A thermal neutrino interaction rate at NLO,}
  Nucl.\ Phys.\ B {950} (2020) 114870
  [1910.12880].
  %%CITATION = doi:10.1016/j.nuclphysb.2019.114870;%%

\bibitem{width5}
  G.D.~Moore, S.~Schlichting, N.~Schlusser and I.~Soudi,
  {\it Non-perturbative determination of collisional broadening
  and medium induced radiation in QCD plasmas,}
  JHEP {10} (2021) 059
  [2105.01679].
  %%CITATION = doi:10.1007/JHEP10(2021)059;%%

\bibitem{screening}
  B.B.~Brandt, A.~Francis, M.~Laine and H.B.~Meyer,
  {\it A relation between screening masses and real-time rates,}
  JHEP {05} (2014) 117
  [1404.2404].
  %%CITATION = doi:10.1007/JHEP05(2014)117;%%

\end{thebibliography}
\end{document}